\documentclass[preprint,aps,tightenlines,showpacs]{revtex4}

\usepackage[dvips]{graphicx}

\usepackage{dcolumn}     

\renewcommand{\case}{\frac}

\begin{document}
 
{  

\title{Quantum Monte Carlo calculations of $A=9,10$ nuclei}

\author
{ Steven C. Pieper$^{1,}$\cite{scp},
  K. Varga$^{1,2,}$\cite{kv},
  R. B. Wiringa$^{1,}$\cite{rbw} }

\affiliation
{$^1$Physics Division, Argonne National Laboratory, Argonne, Illinois 60439 \\
 $^2$Solid State Division, Oak Ridge National Laboratory, Oak Ridge,
         Tennessee 37831 \\}

\date{\today}

\begin{abstract}
We report on quantum Monte Carlo calculations of the ground and low-lying excited
states of $A=9,10$ nuclei using realistic Hamiltonians containing the
Argonne $v_{18}$ two-nucleon potential alone or with one of several
three-nucleon potentials, including Urbana IX and three of the new Illinois
models.
The calculations begin with correlated many-body wave functions that have 
an $\alpha$-like core and multiple p-shell nucleons, $LS$-coupled to 
the appropriate $(J^{\pi};T)$ quantum numbers for the state of interest.
After optimization, these variational trial functions are used as input
to a Green's function Monte Carlo calculation of the energy, using a 
constrained path algorithm.
We find that the Hamiltonians that include Illinois three-nucleon 
potentials reproduce ten states in $^9$Li, $^9$Be, $^{10}$Be, and $^{10}$B 
with an rms deviation as little as 900 keV.
In particular, we obtain the correct 3$^+$ ground state for $^{10}$B, whereas
the Argonne $v_{18}$ alone or with Urbana IX predicts a 1$^+$ ground state.
In addition, we calculate isovector and isotensor energy differences, 
electromagnetic moments, and one- and two-body density distributions.
\end{abstract}
 

\maketitle

}
 
\section{INTRODUCTION}

In a series of papers \cite{PPCW95,PPCPW97,WPCP00}, we have reported
quantum Monte Carlo (QMC) calculations of ground and low-lying excited state 
energies in $A \leq 8$ nuclei for realistic nuclear Hamiltonians.
These calculations employed the Argonne $v_{18}$ (AV18) two-nucleon potential
\cite{WSS95} and the Urbana IX (UIX) three-nucleon potential \cite{PPCW95}, and are
accurate to $\approx 1-2\%$ of the binding energy for light p-shell nuclei.
More recently, we have used the quantum Monte Carlo calculations to help
construct a series of new and improved pion-exchange-based three-nucleon 
potentials, designated the Illinois models \cite{PPWC01}.
The five Illinois models (IL1-IL5), when used in conjunction with AV18, each
reproduce the experimental energies of 17 narrow states in $A \leq 8$ nuclei 
with an rms deviation of $\approx 400$ keV. 
This contrasts with a 2.3 MeV rms deviation for the AV18/UIX Hamiltonian, and 
7.7 MeV for AV18 alone.

In this paper, we report the extension of our QMC calculations to $A=9,10$
nuclei. 
The QMC methods include both variational (VMC) and Green's function Monte
Carlo (GFMC) methods.
The VMC method is used to construct a variational wave function as a product 
of two and three-body correlation operators acting on a independent-particle
wave function that has an $\alpha$-like core and multiple p-shell
nucleons, $LS$-coupled to the appropriate $(J^{\pi};T)$ quantum numbers for
the state of interest.
Monte Carlo evaluation of the energy expectation value is used to optimize
the trial function, particularly the mix of independent-particle wave function
components.
The GFMC method starts from this trial function and makes a Euclidean propagation
that converges to the lowest energy for a state of these quantum numbers.
A constrained path algorithm is crucial for keeping the fermion sign problem
under control.
At present the GFMC method is used to calculate only the lowest state of
given $(J^{\pi};T)$.

We have calculated ten stable or very narrow natural-parity states in the 
$A=9,10$ nuclei $^9$Li, $^9$Be, $^{10}$Be, and $^{10}$B that are experimentally
well known.
We use three of the new Illinois models (IL2, IL3, IL4) in conjunction with 
AV18 and obtain rms deviations from the experimental energies of the states 
of 900, 1100, and 1700 keV, respectively.
We have also calculated most of these states with the AV18 and AV18/UIX 
Hamiltonians for comparison.
The most intriguing result we find is that the new AV18/IL2-IL4 models all
correctly predict a 3$^+$ ground state for $^{10}$B, but the older models 
wrongly predict a 1$^+$ ground state.
In addition, we have made GFMC calculations of six other
states that are expected within the p-shell formulation of these 
nuclei and of the $^9$He and $^{10}$He ground states; these states either have much 
larger widths or are not clearly identified by experiment.

We review briefly the experimental status of the ground
and low-lying excited states in the $A=9,10$ nuclei in Sec.~II.
The Hamiltonians are described in Sec.~III and the QMC calculations in Sec.~IV.
Most of this material has been discussed in detail in
Refs.~\cite{PPCPW97,WPCP00}.
The only major new technical development is the automation of the 
construction of the independent-particle portion of the variational trial 
wave functions that serve as starting points for the GFMC calculations.
Energy results are given in Sec.~V, while electromagnetic moments and
density distributions are shown in Sec.~VI.
We present our conclusions in Sec.~VII.

\section{EXPERIMENTAL STATUS}

\begin{figure}[ht!]
\centering
\includegraphics[width=5.0in]{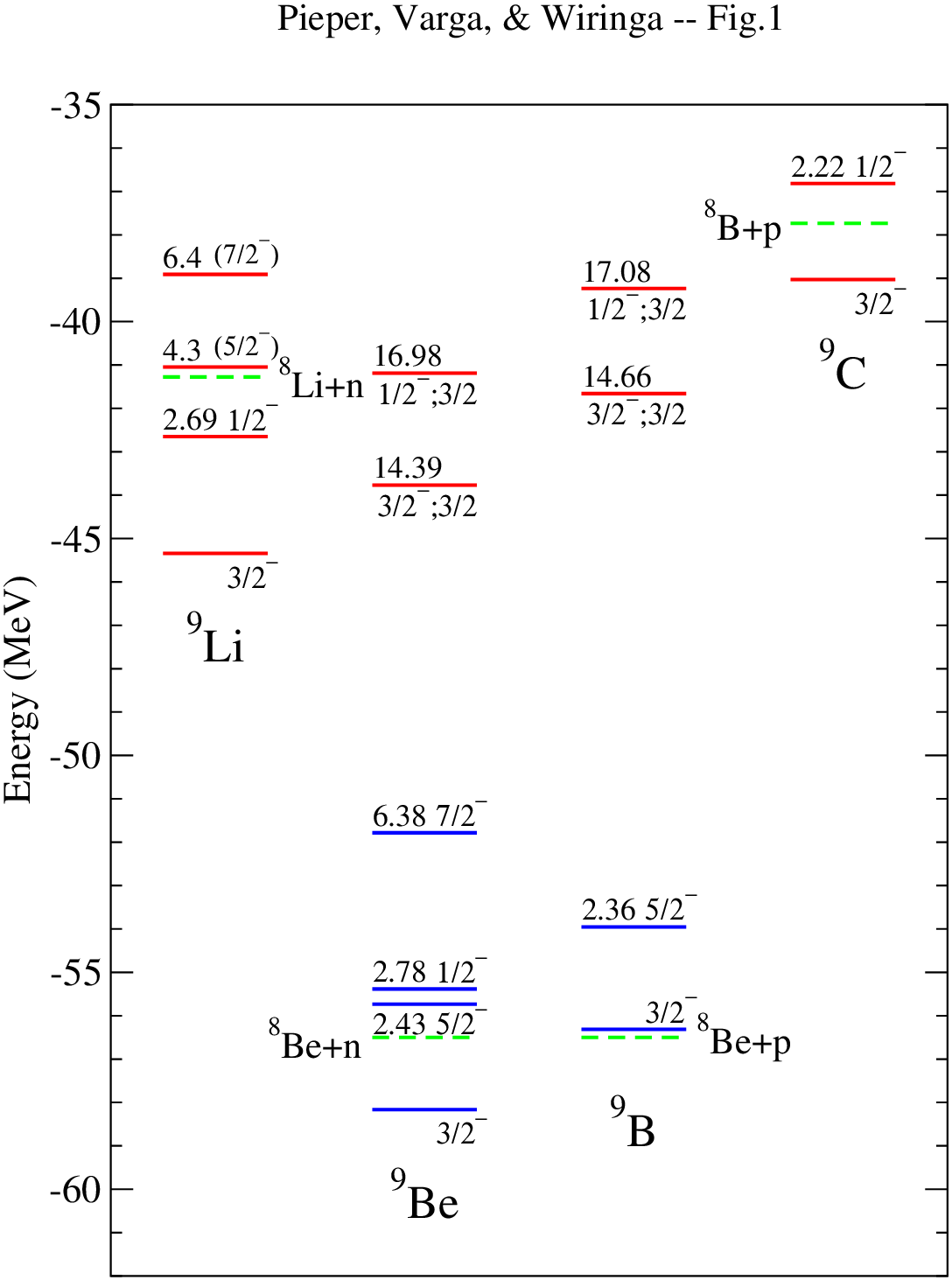}
\caption{First natural-parity states in the experimental spectra of $A=9$ 
nuclei.}
\label{fig:fig1}
\end{figure}

\begin{figure}[ht!]
\centering
\includegraphics[width=5.0in]{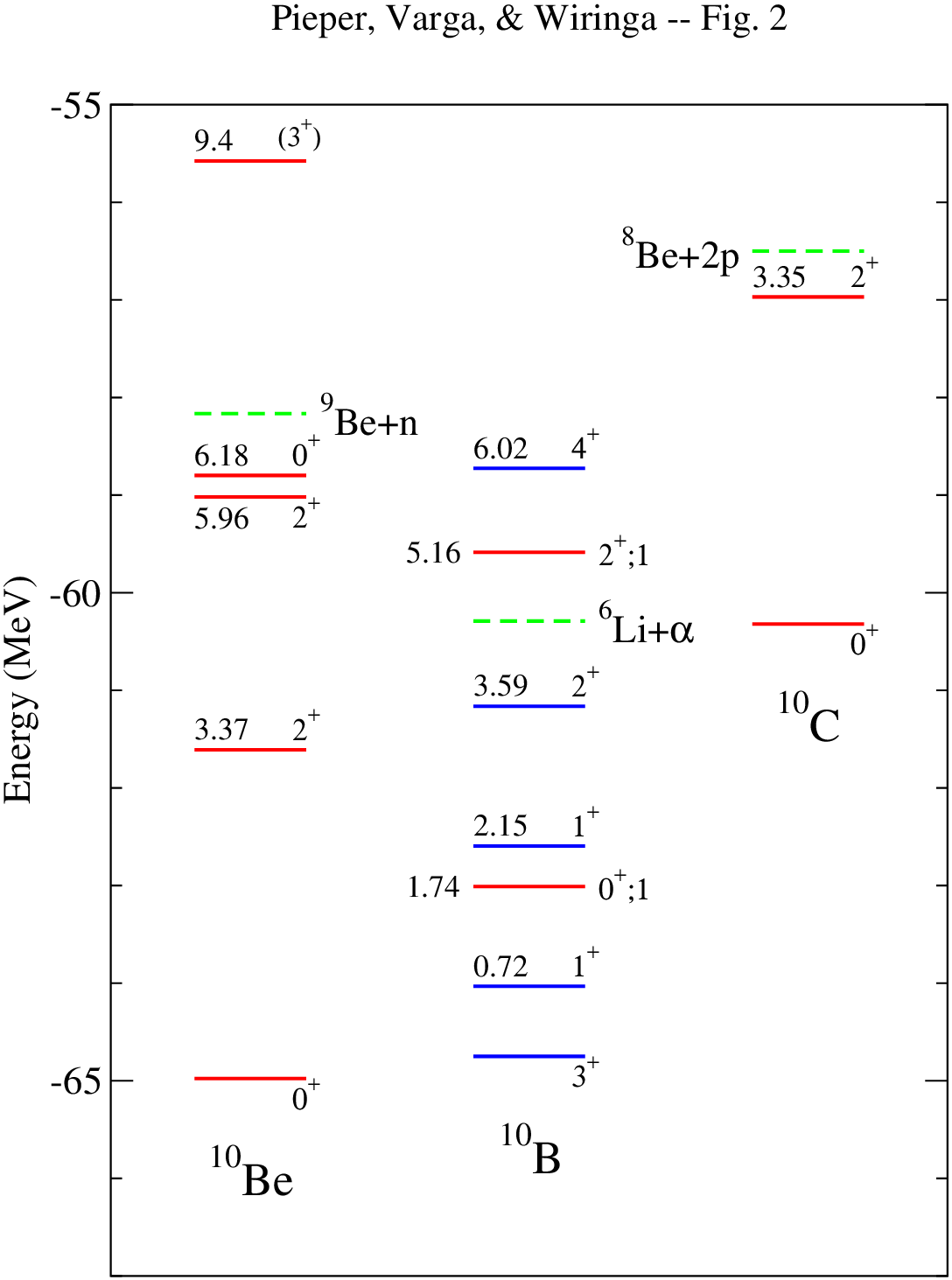}
\caption{Low-lying natural-parity states in the experimental spectra of $A=10$ 
nuclei.}
\label{fig:fig2}
\end{figure}

The experimental status of $A=9,10$ nuclei is illustrated in 
Figs.~\ref{fig:fig1} and \ref{fig:fig2}, where we show the ground states
and most low-lying natural-parity states whose spin assignments are 
reasonably certain~\cite{AS88,Til01}.
We also show some additional narrow states in $^9$Li and $^{10}$Be whose 
spins have not been determined experimentally; reasonable guesses are
given in parentheses, based on standard shell model studies and the
present calculations.

The ground state of $^9$Be is an absolutely stable
$(J^{\pi};T) = (\case{3}{2}^-;\case{1}{2})$ state.
The first excitation is a $(\case{1}{2}^+;\case{1}{2})$ state (not shown in
Fig.~\ref{fig:fig1}) with a width of 217 keV which occurs at the threshold 
for breakup into $^8{\rm Be}+n$.
This unnatural parity state is beyond the scope of the present paper;
we will report calculations of such intruder states in the future.
The second excitation is a narrow $(\case{5}{2}^-;\case{1}{2})$ state
at 2.429 MeV, with a width $<1$ keV.
Within the p-shell formulation, there can also be
$(\case{1}{2}^-;\case{1}{2})$, $(\case{7}{2}^-;\case{1}{2})$, and
$(\case{9}{2}^-;\case{1}{2})$ states.
Experimentally, the first two are observed at 2.78 and 6.38 MeV excitation
but both are quite broad ($\approx 1$ MeV), while no state with
$(\case{9}{2}^-;\case{1}{2})$ character has been identified.
We evaluate all these $(J^{\pi};T)$ cases in GFMC, treating the states as if 
they were particle stable; this should be adequate for narrow states, 
but may be less satisfactory for broad states.
(Several additional unnatural parity states and second excited states of given
$(J^{\pi};T)$ are observed above 3 MeV, but are not shown in the figure.)
The matrix elements of the electromagnetic and strong 
charge-independence-breaking terms in the Hamiltonian are evaluated
perturbatively to infer the energies of the narrow $T=\case{1}{2}$ isobaric 
analog states in $^9$B.

The ground state of $^9$Li is a particle stable ($\case{3}{2}^-;\case{3}{2}$)
state that decays by $\beta^-$ emission to $^9$Be with a half-life of 178 ms.
The first excited state of $^9$Li at 2.691 MeV is believed to be a
($\case{1}{2}^-;\case{3}{2}$) state; it is below the threshold for
breakup into $^8{\rm Li}+n$ and decays only by $\gamma$ emission.
We calculate both these states in $^9$Li, and again perturbatively evaluate 
the energies of the isobaric analogs in $^9$Be, $^9$B, and $^9$C.
Two additional narrow states above the breakup threshold have been observed in
$^9$Li, but without firm spin-parity identification: a 60 keV wide state
at 4.296 MeV and a 40 keV wide state at 6.43 MeV.
Within the p-shell, these should be ($\case{5}{2}^-;\case{3}{2}$) and
($\case{7}{2}^-;\case{3}{2}$) states; our GFMC evaluations of such states line up
very well with the experimental observations, suggesting these
spin assignments may indeed be correct.

The ground state of $^{10}$B is an absolutely stable (3$^+$;0) state.
The first threshold for breakup of $^{10}$B is into $^6$Li+$\alpha$.
Between the ground state and breakup threshold are two (1$^+$;0) excited states
at 0.718 MeV and 2.154 MeV, one (2$^+$;0) state at 3.587 MeV, and the 
(0$^+$;1) isobaric analog of $^{10}$Be at 1.740 MeV.
Many additional states are known above the $^6$Li+$\alpha$ threshold;
in Fig.~\ref{fig:fig2} we show only the (2$^+$;1) isobaric analog at
5.164 MeV and the (4$^+$;0) state at 6.025 MeV.
We calculate the (3$^+$;0) ground state and first (1$^+$;0), (2$^+$;0), and
(4$^+$;0) excitations by GFMC.
In the p-shell there can also be (0$^+$;0) and (5$^+$;0) states, but the
former is of low spatial symmetry while the latter has high angular momentum,
leading us to expect both of them to be quite high in excitation energy; no
experimental observation of either has been made.

The nucleus $^{10}$Be is particle-stable but decays by $\beta^-$ emission with
the very long half-life of $1.51 \times 10^6$ years.
As is typical for an even-even nucleus, the ground state is a (0$^+$;1) state,
and there is a well-separated (2$^+$;1) excited state at 3.368 MeV.
These are the two primary $^{10}$Be states we evaluate in GFMC.
In addition, the tabulation lists a state at 9.4 MeV as a possible (2$^+$;1) 
state, but a more recent (t,$^3$He) experiment makes a (3$^+$;1) assignment 
more likely~\cite{10Be3,Millener}.
We have made one computation for the (3$^+$;1) level using AV18/IL2 and
get reasonable agreement with this energy.
Other particle-stable states below the threshold for breakup into $^9$Be+$n$ 
include second (0$^+$;1) and (2$^+$;1) states, also shown in 
Fig.~\ref{fig:fig2}, and the first particle-stable (1$^-$;1) and (2$^-$;1) 
intruder states (not shown).
The second excited states are not evaluated at present in GFMC, 
while the intruder states will be reported on in future work.
Again, the isobaric analog states in $^{10}$B and $^{10}$C are evaluated 
perturbatively from $^{10}$Be.

In addition to these nuclei, we have also made calculations of the expected
lowest natural-parity states in $^9$He and $^{10}$He.  
These are a $(\case{1}{2}^-;\case{5}{2})$ resonance which is observed at
$\approx 1.2$ MeV above the threshold for breakup into $^8$He+$n$, and a 
$(0^+;3)$ resonance at $\approx 1.1$ MeV above the $^8$He+$n$+$n$ threshold
\cite{Til01,KPB99}.
These resonances are observed to be reasonably narrow, with widths
of 100 and $\leq 300$ keV, respectively.
There are recent experimental reports of a $(\case{1}{2}^+;\case{5}{2})$ 
resonance near threshold in $^9$He~\cite{msu-expt}, which we have not
attempted to calculate; such a state is likely to be very broad and really
needs to be treated as a scattering state.
We also have not attempted to calculate the resonant ground state of 
$^{10}$Li, which is broad and still has some experimental uncertainty.

\section{HAMILTONIAN}

The Hamiltonian includes nonrelativistic one-body kinetic energy, the AV18
two-nucleon potential~\cite{WSS95} and either the UIX~\cite{PPCW95} or 
one of the Illinois~\cite{PPWC01} three-nucleon potentials:
\begin{equation}
   H = \sum_{i} K_{i} + \sum_{i<j} v_{ij} + \sum_{i<j<k} V_{ijk} \ .
\end{equation}
The kinetic energy operator is predominantly charge-independent (CI), but
has a small charge-symmetry breaking (CSB) component due to the difference 
between proton and neutron masses.
The AV18 is one of a class of highly accurate $N\!N$ potentials that fit both 
$pp$ and $np$ scattering data up to 350 MeV with a $\chi^2/$datum $\approx 1$.
It can be written as a sum of electromagnetic and one-pion-exchange terms 
and a shorter-range phenomenological part:
\begin{equation}
   v_{ij} = v^{\gamma}_{ij} + v^{\pi}_{ij} + v^{R}_{ij} \ .
\end{equation}
The electromagnetic terms include one- and two-photon-exchange Coulomb
interaction, vacuum polarization, Darwin-Foldy, and magnetic moment terms,
all with appropriate proton and neutron form factors.
The one-pion-exchange part of the potential includes the small charge-dependent 
(CD) terms due to the difference in neutral and charged pion masses.
The shorter-range part has about 40 parameters which are adjusted to
fit the $pp$ and $np$ scattering data, the deuteron binding energy, and
also the $nn$ scattering length.

The one-pion-exchange and the remaining phenomenological part of the potential
can be written as a sum of eighteen operators,
\begin{equation}
       v^{\pi}_{ij} + v^{R}_{ij} = \sum_{p=1,18} v_{p}(r_{ij}) O^{p}_{ij} \ .
\label{eq:operator}
\end{equation}
The first fourteen CI operators are
\begin{eqnarray}
O^{p=1,14}_{ij} = [1, {\bf\sigma}_{i}\cdot{\bf\sigma}_{j}, S_{ij},
{\bf L\cdot S},{\bf L}^{2},{\bf L}^{2}{\bf\sigma}_{i}\cdot{\bf\sigma}_{j},
({\bf L\cdot S})^{2}]\otimes[1,{\bf\tau}_{i}\cdot{\bf\tau}_{j}] \ ,
\end{eqnarray}
while the last four,
\begin{equation}
   O^{p=15,18}_{ij} = [1, {\bf\sigma}_{i}\cdot{\bf\sigma}_{j},
S_{ij}]\otimes T_{ij} , \ \tau_{zi}+\tau_{zj} \ ,
\end{equation}
are the strong interaction CD and CSB terms.

The three-nucleon potentials from the Urbana series of models~\cite{CPW83}
contain a long-range, two-pion-exchange, P-wave term and a short-range 
phenomenological piece:
\begin{equation}
   V^{U}_{ijk} = V^{2\pi,P}_{ijk} + V^{R}_{ijk} \ .
\end{equation}
The UIX model has the strengths of these two terms adjusted to reproduce the 
binding energy of $^3$H, in GFMC calculations, and to give a reasonable 
saturation density in nuclear matter, in variational chain summation 
calculations~\cite{AP97}, when used with AV18.
The Illinois models add a two-pion-exchange S-wave term and a three-pion-ring 
term:
\begin{equation}
   V^{IL}_{ijk} = V^{2\pi,P}_{ijk} + V^{2\pi,S}_{ijk} + V^{3\pi,\Delta R}_{ijk} 
                + V^{R}_{ijk} \ .
\end{equation}
The two-pion-exchange S-wave term is required by chiral symmetry, but in 
practice its small energy contribution makes it hard to distinguish from the 
dominant P-wave term.
However, the three-pion-ring term, while it is smaller than the
two-pion-exchange P-wave term, has a distinctly different isospin dependence,
which is crucial for being able to fit the variety of light p-shell energy
levels studied in Ref.~\cite{PPWC01}.
In the Illinois models, the operator structure and radial forms were taken
from standard meson-exchange theory, but the overall strengths of the four
terms, and one cutoff factor in the radial dependence, were adjusted
to obtain best fits to the energies of 17 narrow states in $3\leq A\leq 8$
nuclei.
In practice, at most three parameters at a time could be uniquely determined 
from the energy calculations, so five different models were
constructed in which different subsets of the parameters were fixed
by external considerations, while the remaining ones were adjusted.

The CD and CSB terms in $H$ are fairly weak, so we can treat them
conveniently as a first-order perturbation and use a wave function
of good isospin, which is significantly more compact.
Also, direct GFMC calculations with the spin-dependent terms that
involve the square of the orbital angular momentum operator can have large 
statistical fluctuations, as discussed in Ref.~\cite{PPCPW97}.
Thus it is useful to define~\cite{PPCPW97} a simpler isoscalar interaction, 
AV8$^\prime$, which contains only the eight operators
$[1, {\bf\sigma}_{i}\cdot{\bf\sigma}_{j}, S_{ij},{\bf L\cdot S}]
\otimes[1,{\bf\tau}_{i}\cdot{\bf\tau}_{j}]$ 
and an isoscalar Coulomb interaction. 
These eight operators are chosen such that AV8$^\prime$ reproduces the
CI part of the full AV18 interaction in all S and P waves as well
as the deuteron.
The AV8$^\prime$ interaction 
(without the Coulomb term) was recently used in a benchmark test of seven 
different many-body methods for solving the four-nucleon bound state,
with excellent agreement between GFMC and the other calculations~\cite{KNG01}.

\section{QUANTUM MONTE CARLO}

\subsection{Variational Monte Carlo}

We first construct a variational wave function for the state of interest
and then optimize it by minimizing the energy expectation value as computed
by Metropolis Monte Carlo integration.
The variational wave function for the nuclei studied here has the form
\begin{eqnarray}
|\Psi_{V}\rangle =  \Big[1 + \sum_{i<j<k}\tilde{U}^{TNI}_{ijk} \Big]
                   {\cal S} \prod_{i<j} \Big( 1+U_{ij} \Big) |\Psi_J\rangle \ .
\label{eq:psiv}
\end{eqnarray}
The $U_{ij}$, and $\tilde{U}^{TNI}_{ij;k}$ are noncommuting two- and 
three-nucleon correlation operators, and ${\cal S}$ is a symmetrization 
operator.
The $U_{ij}$ includes spin, isospin, and tensor terms induced by the 
two-nucleon potential, while the $\tilde{U}^{TNI}_{ij;k}$ reflects the 
structure of the dominant parts of the three-nucleon interaction.
This trial function has the advantage of being efficient to evaluate while
including the bulk of the correlation effects.
A more sophisticated variational function can be constructed by including
two-body spin-orbit correlations and additional three-body correlations,
as discussed in Ref.~\cite{WPCP00}, but the time to compute these extra 
terms is significant, while the gain in the variational energy is relatively
small.
Studies have shown that the GFMC algorithm easily corrects for the omission
of these terms~\cite{PPCPW97}.

The two-body correlations are generated by the solution of coupled 
differential equations with embedded variational parameters \cite{W91}.
We have found that the parameters optimized for the $\alpha$-particle are 
near optimal for use in the light p-shell nuclei.
Likewise, the best parameters for the three-body correlations are
remarkably constant for different s- and p-shell nuclei, so they have
not been changed significantly from the previous $A \leq 8$ work 
\cite{PPCPW97,WPCP00}.

For the p-shell nuclei studied here, the totally antisymmetric Jastrow wave 
function, $\Psi_J$, starts with a sum over independent-particle terms, 
$\Phi_A$, that have 4 nucleons in an $\alpha$-like core and $(A-4)$ nucleons
in p-shell orbitals.
We use $LS$ coupling to obtain the desired $JM$ value of a given state, 
as suggested by standard shell-model studies~\cite{CK65}.
We also need to specify the spatial symmetry $[n]$ of the angular
momentum coupling of the p-shell nucleons~\cite{BM69}.
Different possible $LS[n]$ combinations lead to multiple components in the
Jastrow wave function.
This independent-particle basis is acted on by products of central pair 
and triplet correlation functions, which depend upon the shells (s or p) 
occupied by the particles and on the $LS[n]$ coupling:
\begin{eqnarray}
  |\Psi_{J}\rangle &=& {\cal A} \left. \Big\{ \right.
     \prod_{i<j<k \leq 4} f^{c}_{ijk} 
     \prod_{i<j \leq 4}f_{ss}(r_{ij})
     \prod_{k \leq 4<l \leq A} f_{sp}(r_{kl})  \nonumber\\
                   && \sum_{LS[n]}
     \Big( \beta_{LS[n]} \prod_{4<l<m \leq A} f^{[n]}_{pp}(r_{lm}) \ 
    |\Phi_{A}(LS[n]JMTT_{3})_{1234:5\ldots A}\rangle \Big) \left. \right. \Big\} \ .
\label{eq:jastrow}
\end{eqnarray}
The operator ${\cal A}$ indicates an antisymmetric sum over all possible
partitions of the $A$ particles into 4 s-shell and $(A-4)$ p-shell ones.
The pair correlation for particles within the s-shell, $f_{ss}(r)$, 
is the optimal correlation for the $\alpha$-particle.
The $f_{sp}(r)$ is similar to the $f_{ss}(r)$ at short range, but with a 
long-range tail going to a constant $\approx 1$; this allows the wave function 
to develop a cluster structure like $\alpha + d$ in $^6$Li or 
$\alpha + \alpha$ in $^8$Be at large cluster separations.
The $f^{[n]}_{pp}(r)$ depends on the nucleus and particular independent-particle
channel, e.g., in the case of $^6$Li or $^8$Be, it is similar to the optimal
deuteron or alpha correlations.

The $LS[n]$ components of the independent-particle wave function are given by:
\begin{eqnarray}
 &&  |\Phi_{A}(LS[n]JMTT_{3})_{1234:5\ldots A}\rangle =
     |\Phi_{\alpha}(0 0 0 0)_{1234} \prod_{4 < l\leq A}
     \phi^{LS[n]}_{p}(R_{\alpha l}) \nonumber \\
 &&  \left\{ [ \prod_{4 < l\leq A} Y_{1m_l}(\Omega_{\alpha l}) ]_{LM_L[n]}
     \otimes [ \prod_{4 < l\leq A} \chi_{l}(\case{1}{2}m_s) ]_{SM_S}
     \right\}_{JM}
     \otimes [ \prod_{4 < l\leq A} \nu_{l}(\case{1}{2}t_3) ]_{TT_3}\rangle \ , 
\label{eq:phia}
\end{eqnarray}
where
\begin{equation}
     \Phi_{\alpha}(0 0 0 0)
     = {\cal A} ( p\uparrow p\downarrow n\uparrow n\downarrow )
\end{equation}
is the $\alpha$-core independent-particle wave function.
The $\phi^{LS}_{p}(R_{\alpha l})$ are $p$-wave solutions of a particle 
of reduced mass $\case{4}{5}m_n$ in an effective $\alpha$-$N$ potential:
\begin{equation}
   V_{\alpha N}(r) = V^{WS}_{\alpha N}(r) + V^C_{\alpha N}(r) \ .
\end{equation}
The $\phi^{LS}_{p}$ are functions of the distance between the center of mass 
of the $\alpha$ core (which contains particles 1-4 in this partition)
and nucleon $l$, and again may be different for different $LS[n]$ 
components.
For each state considered in the present work, we have used
bound-state asymptotic conditions for the $\phi^{LS}_{p}$, even if
the state is particle unstable.
The Woods-Saxon potential
\begin{equation}
   V^{WS}_{\alpha N}(r) = V^{LS}_p [1+{\rm exp}(\frac{r-R_p}{a_p})]^{-1} \ ,
\label{eq:spwell}
\end{equation}
has variational parameters $V^{LS}_p$, $R_p$, and $a_p$, while the Coulomb 
potential is obtained by folding over nuclear and nucleon form factors.
The wave function is translationally invariant, hence there is no 
spurious center of mass motion.

A major technical advance in the present work is the automatic generation of
the independent-particle wave function $\Phi_A$ with the appropriate spatial 
symmetries discussed above.
We describe here the construction of 
the spatial symmetry $[n]$.
One can use different coupling schemes to form the spatial 
(or spin and isospin) part of the wave function with good 
quantum numbers. We use simple sequential coupling in which
the spatial part of the wave function is
written as
\begin{equation}
\theta_{LM_L\mu}=\left[\left[\left[
Y_{1}({\Omega_{\alpha 5}})
Y_{1}({\Omega_{\alpha 6}})\right]_{l_{56}} 
Y_{1}({\Omega_{\alpha 7}})\right]_{l_{567}}....\right]_{LM_L} ,
\end{equation}
and a similar construction is used for the spin and isospin part. 
The functions having the same $L$ and $M_L$ but different 
intermediate quantum numbers, labeled by $\mu$,  
are orthogonal and form a complete set of eigenstates of 
$L^2$ and $L_z$. The permutation operators $P_k$ 
of the valence particles  $l=5,6,...,A$ ($k=1,...,(A-4)!$), 
commute with  $L^2$ and $L_z$ so that the above 
functions form a basis for the representation of the symmetric group as well:
\begin{equation}
P_k\theta_{LM_L\mu}=\sum_{\lambda} U_{\lambda\mu}(P_k)\theta_{LM_L\lambda} \ .
\end{equation}

The permutation symmetry is conveniently depicted by using ``Young 
diagrams'', consisting of $N$ adjoining square boxes with rows 
numbered numerically downward, and columns rightward; there may not 
be more rows in column $i$ than in column $i-1$, nor columns in row 
$j$ than in row $j-1$. Each Young diagram corresponds to a representation
of the permutation group. The basis functions defining a given
representation can be labeled by using a Young tableau, which is an 
arrangement of the numbers $1,2,...,N$ in the Young diagram, such that 
numbers always increase along all rows and down all columns. 

We use the ``Young operators'' to construct a basis function that has the 
symmetry properties of a given Young tableau.
The Young operator, ${\hat Y}$, is 
a product of a symmetrizer $R$ and an antisymmetrizer $Q$: ${\hat Y}=QR$. The 
operator $R$ symmetrizes all particle indices which are in the same row 
and $Q$ antisymmetrizes all particles in the same column. Both operators
are constructed as a combination of the permutation operators $P_k$.

To construct the spatial part of the wave function belonging to a given
representation of the permutation group, we first have to draw the
Young diagram and insert the numbers $1,2,...,N$ into the pattern in any
order to give a Young tableau. The number of different tableaux $N_Y$ 
is the dimension of the representation. Then we prepare the Young 
operators ${\hat Y}_1,...,{\hat Y}_{N_Y}$ corresponding to the Young tableau and 
calculate the matrix elements of these operators with the basis functions
$\theta_{LM_L\mu}$. We then have
\begin{equation}
{\hat Y}_k \theta_{LM_L\mu}=\sum_{\lambda} 
\langle\theta_{LM_L\mu}\vert {\hat Y}_k\vert
\theta_{LM_L\lambda}\rangle \theta_{LM_L\lambda} .
\end{equation}
Let us denote the successively coupled spin  
(isospin) functions by $\omega_{SM_Snu}$ ($\omega_{TT_3\nu'}$).
To form an antisymmetric wave function for the $N$ particle system
one has to multiply the basis functions of the space part of a 
given Young tableau, ${\hat Y}_k\theta_{LM_L\mu}$, by the basis functions of
the spin-isospin part belonging to the conjugated $[{\bar n}]$ Young 
diagram (obtained by reversal of the roles of rows and columns) 
${\hat Y}_k[{\bar n}]\omega_{SM_S\nu}\omega_{TT_3\nu'}$ and sum 
over all possible tableaux.
Thus, Eq.(\ref{eq:phia}) becomes:
\begin{eqnarray}
     |\Phi_{A}(LS[n]JMTT_{3})_{1234:5\ldots A}\rangle &=&
     |\Phi_{\alpha}(0 0 0 0)_{1234} \prod_{4 < l\leq A}
     \phi^{LS[n]}_{p}(R_{\alpha l}) \nonumber \\
 && \sum_k p_k {\hat Y}_k[n] {\hat Y}_k[{\bar n}]
\left[\theta_{L\mu} \omega_{S\nu}\right]_{JM}
\omega_{TT_3\nu'} \rangle \ ,
\end{eqnarray}
where $p_k$ is the parity of the permutation
of the numbers (starting from the top and going 
left to right in each row) in the Young diagram. 
The Young operator ${\hat Y}_k[n]$ (${\hat Y}_k[{\bar n}]$) acts
on the spatial (spin-isospin) functions of the $l=5,6,...,A$ valence particles.
Any choice of $\mu$ and $\nu,\nu'$ generates the same wave function. 

The different possible $LS[n]$ contributions to $A=9,10$ nuclei are given in
Table~\ref{tab:perms}.
After other parameters in the trial function have been optimized, a series of 
energy evaluations are made in which the $\beta_{LS[n]}$ of 
Eq.(\ref{eq:jastrow}) are different in the left- and right-hand-side 
wave functions to obtain the diagonal and off-diagonal matrix elements of 
the Hamiltonian and the corresponding normalizations and overlaps.
The resulting N$\times$N matrices are diagonalized to find the $\beta_{LS[n]}$
eigenvectors, using generalized eigenvalue routines because the correlated
$\Psi_V$ are not orthogonal.
This allows us to project out not only the ground-state trial functions, but
also excited-state trial functions of the same $(J^\pi;T)$ quantum numbers.
In our present studies, we have carried out our $A$=9 diagonalization in a
complete p-shell basis, but for $^{10}$B and $^{10}$Be we have limited 
ourselves to the three highest spatial symmetries, i.e., [42], [33], and [411].
The diagonalization is carried out for the AV18/UIX Hamiltonian; the 
$\beta_{LS[n]}$ amplitudes should not be significantly different for the other
models.
Additional spatial symmetries involving particle excitations out of the
p-shell are built up in the full trial function by means of the tensor 
correlations contained in the $U_{ij}$ and $\tilde{U}^{TNI}_{ijk}$ of 
Eq.(\ref{eq:psiv}).

Thus in $^9$Be, the Jastrow wave function for the $(\case{3}{2}^-;\case{1}{2})$
ground state is constructed from thirteen amplitudes, and a 
$13 \times 13$ diagonalization is performed to find the optimal mixing.
The $\beta_{LS[n]}$ values for this and other $^9$Be states, including some
second excited states, are given in Table~\ref{tab:be9}, while the
values for various states in $^9$Li are given in Table~\ref{tab:li9}.
In the case of $^9$Be, the [41] symmetry states dominate; addition
of the lower spatial symmetries improves the energies by typically 0.25 MeV.
The additional states do not significantly alter the rms radii or
electromagnetic moments.
By contrast, the leading [32] symmetry in $^9$Li is not so dominant;
addition of lower spatial symmetries gives a significant 1--2 MeV improvement 
to some of the energies.
Electromagnetic moments can also be significantly shifted.
In general, the dominant $A=9$ amplitudes are in good agreement (modulo sign)
with the shell-model wave functions of Kumar~\cite{K74}.

The $\beta_{LS[n]}$ values for $^{10}$B states are given in Table~\ref{tab:b10},
and for $^{10}$Be states in Table~\ref{tab:be10}.
The neglect of the [321] or lower symmetry states in these nuclei is justified on 
the grounds that this is the leading spatial symmetry of $^{10}$Li, whose
isobaric analog state first appears at 21 MeV in the excitation spectrum of
$^{10}$Be.
For $^{10}$B, the [42] symmetry states are dominant, and addition of the
[33] and [411] symmetries improves the energies by only 0.2 MeV.
However, for $^{10}$Be, these extra components can give significant additional 
binding of up to 3 MeV.
In this case, the extra states correspond to low-lying excitations that
may not be filtered out of the trial function by a GFMC propagation to
limited $\tau$, as discussed below.
Thus it is crucial to carry out the diagonalization in the trial function
to get an optimal starting point for the GFMC calculation.

The $A=10$ nuclei are exactly midway through the p-shell and are unique
in having two linearly independent states of the same spatial symmetry 
contribute, i.e., two $^1$D[42] states in $^{10}$Be and two $^3$D[42] 
states in $^{10}$B.
To uniquely identify these two possible combinations, we diagonalize
Jastrow trial functions containing just the two $^{2S+1}$D[42] states
in the quadrupole moment operator, so the first
$\beta_{LS[n]}$ amplitude reported in Tables~\ref{tab:b10}--\ref{tab:be10} 
for each state corresponds to the lower (negative) quadrupole eigenvalue 
and the second to the higher (positive) quadrupole eigenvalue.
Interestingly, we see that most states where these two symmetries contribute
are dominated by either one amplitude or the other, e.g., the first 3$^+$ in 
$^{10}$B is almost pure positive quadrupole while the second 3$^+$
is almost pure negative quadrupole in composition.
Similarly the first 2$^+$ in $^{10}$Be is almost pure negative quadrupole
while the second is pure positive quadrupole.
This is analogous to using the $LK_L$ labeling scheme applied to $^{10}$B by
Kurath in a traditional shell model calculation~\cite{K79}, and there is
good agreement as to the dominant [42] symmetry amplitudes with his 
Table~3.
Only the first 2$^+$ state in $^{10}$B does not seem to fit in with these
observations.

Energy expectation values with the full $\Psi_V$ of Eq.(\ref{eq:psiv}) are 
evaluated by a Metropolis Monte Carlo algorithm as described in \cite{PPCPW97}.
The full wave function at any given spatial configuration 
${\bf R}={\bf r}_1,{\bf r}_2,...{\bf r}_A$
can be represented by a vector of $2^A \times I(A,T)$ complex numbers,
\begin{equation}
  \Psi({\bf R}) = \sum_{\alpha} \psi_{\alpha}({\bf R}) |\alpha\rangle \ ,
\label{eq:psivec}
\end{equation}
where the $\psi_{\alpha}({\bf R})$ are the complex coefficients of each state
$|\alpha\rangle$ which has specific third components of spin and linear
combinations of good isospin.
For the nuclei considered here, this leads to vectors ranging in length from
13,824 for $^9$He to 92,160 for $^{10}$Be, although a savings of a factor of
two is possible in even-$A$ nuclei by computing in the $M=0$ substates.
The spin, isospin, and tensor operators $O^{p=2,6}_{ij}$ contained in the 
potential and other operators of interest are sparse matrices in this basis
and thus easily evaluated.
Kinetic energy and spin-orbit operators require the computation of first
derivatives and diagonal second derivatives of the wave function.
These are obtained by evaluating the wave function at $6A$ slightly shifted
positions of the coordinates ${\bf R}$ and taking finite differences.
Terms quadratic in ${\bf L}$ require mixed second derivatives and
additional wave-function evaluations and finite differences.

\subsection{Green's Function Monte Carlo}

The GFMC method~\cite{C87,C88} projects out the exact lowest-energy state,
$\Psi_{0}$, for a given set of quantum numbers, using
$\Psi_0 = \lim_{\tau \rightarrow \infty} \exp [ - ( H - E_0) \tau ] \Psi_T$,
where $\Psi_{T}$ is an optimized trial function from the VMC calculation.
If the maximum $\tau$ actually used is large enough,
the eigenvalue $E_{0}$ is calculated exactly while other expectation values
are generally calculated neglecting terms of order $|\Psi_{0}-\Psi_{T}|^{2}$
and higher~\cite{PPCPW97}.
In contrast, the error in the variational energy is of order
$|\Psi_{0}-\Psi_{T}|^{2}$, and other expectation values calculated with
$\Psi_{T}$ have errors of order $|\Psi_{0}-\Psi_{T}|$.
In the following we present a brief overview of the nuclear GFMC
method; much more detail may be found in Refs.~\cite{PPCPW97,WPCP00}.

We start with a $\Psi_{T}$ defined as (see Eq.(3.13) of Ref.~\cite{WPCP00})
\begin{eqnarray}
|\Psi_{T}\rangle =  {\cal S} \prod_{i<j} \Big( 1+U_{ij} +
\sum_{k \neq i,j}\tilde{U}^{TNI}_{ij;k}   \Big)
|\Psi_J\rangle \ ,
\label{eq:psitgfmc}
\end{eqnarray}
and define the propagated
wave function $\Psi(\tau)$
\begin{eqnarray}
   \Psi(\tau) = e^{-({H}-E_{0})\tau} \Psi_{T}
              = \left[e^{-({H}-E_{0})\triangle\tau}\right]^{n} \Psi_{T} \ ,
\end{eqnarray}
where we have introduced a small time step, $\triangle\tau=\tau/n$;
obviously $\Psi(\tau=0) =  \Psi_{T}$ and
$\Psi(\tau \rightarrow \infty) = \Psi_{0}$.
The $\Psi (\tau)$ is represented by a vector function of $\bf R$ using
Eq.(\ref{eq:psivec}), and the Green's function,
$G_{\alpha\beta}({\bf R},{\bf R}^{\prime})$ is a matrix function of
$\bf R$ and ${\bf R}^{\prime}$ in spin-isospin space, defined as
\begin{equation}
G_{\alpha\beta}({\bf R},{\bf R}^{\prime})= \langle {\bf R},
\alpha|e^{-({H}-E_{0})\triangle\tau}|{\bf R}^{\prime},\beta\rangle \ .
\label{eq:gfunction}
\end{equation}
It is calculated with leading errors of order $(\triangle\tau)^{3}$.
Omitting the spin-isospin indices $\alpha$, $\beta$ for brevity,
$\Psi({\bf R}_{n},\tau)$ is given by
\begin{equation}
\Psi({\bf R}_{n},\tau) = \int G({\bf R}_{n},{\bf R}_{n-1})\cdots G({\bf R}_{1},
{\bf R}_{0})\Psi_{T}({\bf R}_{0})d{\bf R}_{n-1}\cdots d{\bf R}_{1}d{\bf R}_{0}
 \ ,
\label{eq:gfmcpsi}
\end{equation}
with the integral being evaluated stochastically.
The short-time propagator is constructed with the exact two-body propagator
and additional terms coming from the three-body interaction.

Quantities of interest are evaluated in terms of a ``mixed'' expectation value
between $\Psi_T$ and $\Psi(\tau)$:
\begin{eqnarray}
\langle O \rangle_{\rm Mixed} & = & \frac{\langle \Psi_{T} | O |
\Psi(\tau)\rangle}{\langle \Psi_{T} | \Psi(\tau)\rangle} \nonumber \\
& = & \frac{ \int d {\bf P}_n
\Psi_{T}^{\dagger}({\bf R}_{n}) O G({\bf R}_{n},{\bf R}_{n-1})
\cdots G({\bf R}_{1},{\bf R}_{0})\Psi_{T}({\bf R}_{0})}
{\int d{\bf P}_{n} \Psi_{T}^{\dagger}
({\bf R}_{n})G({\bf R}_{n},{\bf R}_{n-1}) \cdots G({\bf R}_{1},
{\bf R}_{0})\Psi_{T}({\bf R}_{0})}~,
\label{eq:expectation}
\end{eqnarray}
where ${\bf P}_{n} = {\bf R}_{0},{\bf R}_{1},\cdots,{\bf R}_{n}$ denotes the
`path', and
$d{\bf P}_{n} = d{\bf R}_{0} d{\bf R}_{1}\cdots d{\bf R}_{n}$ with
the integral over the paths being carried out stochastically.
The desired expectation values would, of course, have $\Psi(\tau)$ on both
sides; by writing
$\Psi(\tau) = \Psi_{T} + \delta\Psi(\tau)$
and neglecting terms of order $[\delta\Psi(\tau)]^2$, we obtain the approximate
expression
\begin{eqnarray}
\langle O (\tau)\rangle =
\frac{\langle\Psi(\tau)| O |\Psi(\tau)\rangle}
{\langle\Psi(\tau)|\Psi(\tau)\rangle}
\approx \langle O (\tau)\rangle_{\rm Mixed}
     + [\langle O (\tau)\rangle_{\rm Mixed} - \langle O \rangle_T] ~,
\label{eq:pc_gfmc}
\end{eqnarray}
where $\langle O \rangle_T$ is the variational expectation value.

A special case is the expectation value of the Hamiltonian.
The $\langle{H}(\tau)\rangle_{\rm Mixed}$ can be re-expressed as
\begin{eqnarray}
\langle{H}(\tau)\rangle_{\rm Mixed} = \frac{\langle \Psi_{T} |
e^{-({H}-E_{0})\tau /2}{H} e^{-({H}-E_{0})\tau /2} |
\Psi_{T}\rangle}{\langle \Psi_{T} |e^{-({H}-E_{0})\tau /2}
e^{-({H}-E_{0})\tau /2}| \Psi_{T}\rangle} \geq E_{0}~,
\end{eqnarray}
since the propagator $\exp [ - (H - E_0) \tau ] $ commutes with the
Hamiltonian.
Thus $\langle{H}(\tau)\rangle_{\rm Mixed}$ approaches $E_{0}$ in the limit
$\tau\rightarrow\infty$, and the expectation value obeys the variational
principle for all $\tau$.

As mentioned above, the propagation is actually made with a simplified 
Hamiltonian $H^{\prime}$ that includes the CI kinetic energy operator, 
the AV8$^\prime$ two-nucleon interaction including isoscalar Coulomb, 
and a slightly altered $V^{\prime}_{ijk}$ three-nucleon interaction.
The AV8$^\prime$ is a little more attractive than AV18, so a slightly more
repulsive $V^{\prime}_{ijk}$ is used to keep
$\langle H^{\prime} \rangle \approx \langle H \rangle$.
This ensures the GFMC algorithm will not propagate to excessively large
densities due to overbinding.
Consequently, the upper bound property applies to
$\langle{H}^{\prime}(\tau)\rangle$, and $\langle H-H^{\prime} \rangle$ must
be evaluated perturbatively.

Another complication that arises in the GFMC algorithm is the ``fermion sign
problem''.
This arises from the stochastic evaluation of the matrix elements
in Eq.(\ref{eq:expectation}).  
The Monte Carlo techniques used to calculate the path integrals 
leading to $\Psi({\bf R}_n,\tau)$ involve
only local properties, while antisymmetry is a global property.
Thus the propagation can mix in the boson solution.  This has a (much) lower
energy than the fermion solution and thus is exponentially amplified
in subsequent propagations.  In the final integration with the antisymmetric
$\Psi_T$, the desired fermionic part is projected out, but in the
presence of large statistical errors that grow exponentially with $\tau$.
Because the number of pairs that can be exchanged grows with $A$, the
sign problem also grows exponentially with increasing $A$.
For $A{\geq}8$, the errors grow so fast that convergence in $\tau$ cannot
be achieved.

To remedy this situation, a ``constrained path'' approximation has been
developed~\cite{WPCP00}.
The basic idea of the constrained-path method is to discard those
configurations that, in future generations, will contribute only noise to
expectation values.
Many tests of the constrained path have been made and it usually gives
results that are consistent with unconstrained propagation, within statistical
errors, although there are cases in which it converges to the wrong 
energy~\cite{WPCP00}.  
Up to now, this problem was always solved by using
a small number, $n_u=10$ to 20, of unconstrained steps
before evaluating expectation values.
These few unconstrained steps, out of 400 or more total steps,
were enough to damp out errors introduced by the constraint
in all the tests that we had done.
The statistical errors in calculations with $n_u=10$ are not substantially
greater than for $n_u=0$. 
However for $A=10$, the error from $n_u=20$ is 60--90\% larger than that
from $n_u=10$, i.e., 2.5--3.5 more configurations must be used
to get the same error.  In the present work, we find that $A=9,10$
calculations with the AV18/IL3 Hamiltonian (which has the strongest
long-range part of $V_{ijk}$) are particularly sensitive to $n_u$,
and most of these are made with $n_u=20$.  
Tests made with the other Hamiltonians show that $n_u=10$ is enough,
except for $^{9,10}$He.

Previously~\cite{WPCP00} we had found that obtaining reliable results
for $^8$He was particularly difficult.  In the present work we find that
this persists for $^{9,10}$He which show large statistical fluctuations.
In addition the constrained-path method appears to be less reliable than
for other nuclei.  For example, we made a calculation for $^{10}$He with
AV18/IL2 and a $\Psi_T$ that contained no $N\!N\!N$ correlation.  Even
though $n_u=20$ was used, the result was overbound by 1.5(6)~MeV.
(In Ref.~\cite{WPCP00} a calculation of $^6$Li with a much worse $\Psi_T$, 
which contained no $NN$ tensor correlations,
also gave an overbound result but this was corrected with only $n_u=10$.)
The statistical errors using $n_u=40$ are very large, making a 
confirmation of the $n_u=20$ results presented here difficult;
we have attempted this for AV18/IL2 and obtain no significant change in the
binding energy within a statistical error of 0.7 MeV.

\begin{figure}[ht!]
\centering
\includegraphics[width=5.0in]{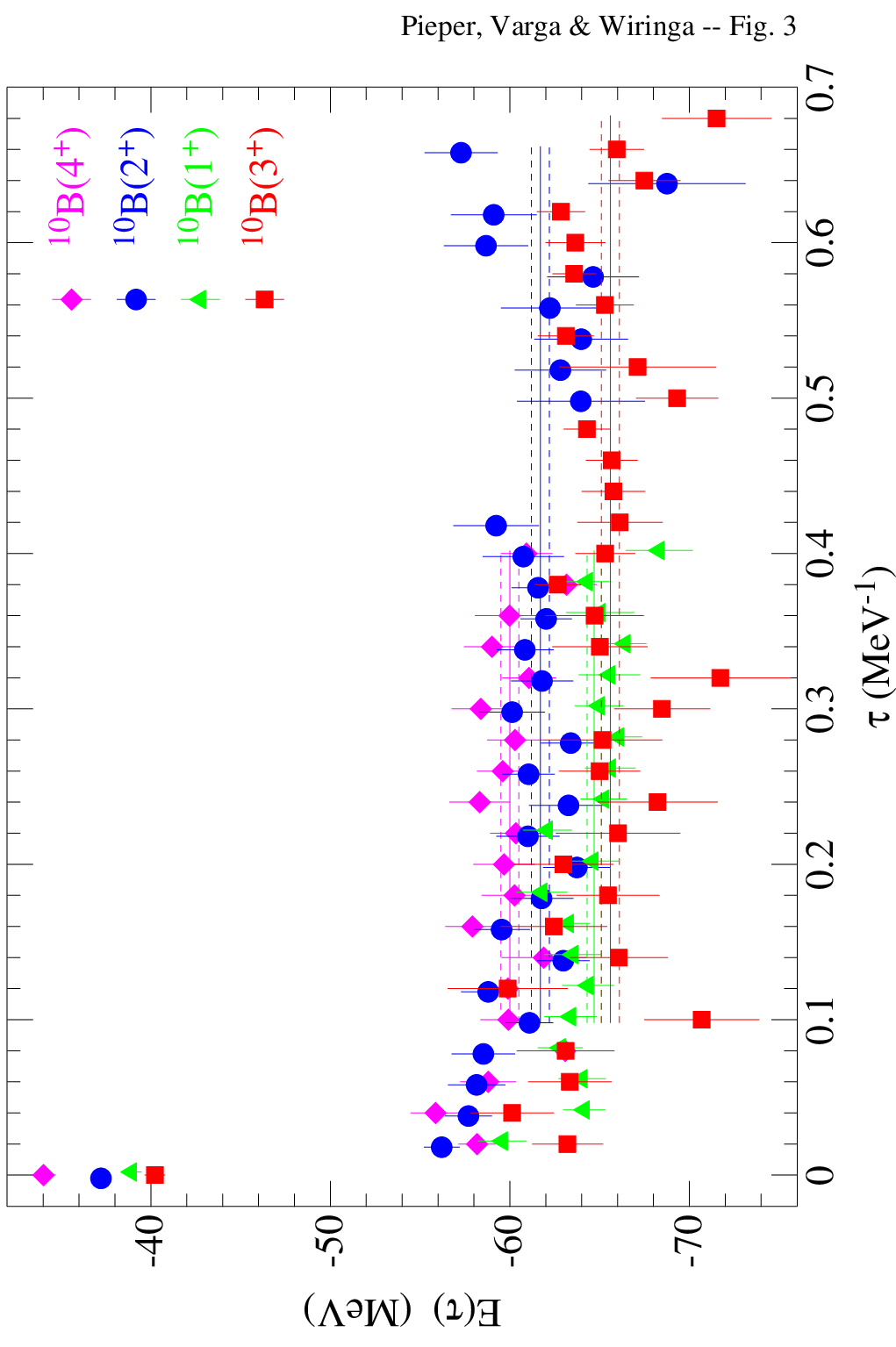}
\caption{GFMC $E(\tau)$ for states of $^{10}$B calculated for the AV18/IL2
Hamiltonian.  The solid lines show the averages that are reported in
the text; the 
dashed lines show the corresponding statistical errors.}
\label{fig:10B-e_of_tau}
\end{figure}

Figure \ref{fig:10B-e_of_tau} shows the progress with increasing $\tau$
of typical constrained GFMC calculations, in this case for various
states of $^{10}$B using the AV18/IL2 Hamiltonian.  The values shown at
$\tau=0$ are the VMC values using $\Psi_T$.  The GFMC very rapidly makes
a large improvement on these energies; by $\tau=0.01$~MeV$^{-1}$, the
$\Psi_T$ energies have been reduced by $\sim$20~MeV.  This rapid
improvement corresponds to the removal of small admixtures of states
with excitation energies $\sim 1$~GeV from $\Psi_T$.  Typically, averages over the
$\tau \geq 0.1$~MeV$^{-1}$ values are used as the GFMC energy.  The
standard deviation, computed using block averaging, of all of the
individual energies for these $\tau$ values is used to compute the
corresponding statistical error.  The solid lines show these averages;
the corresponding dashed lines show the statistical errors.

\section{ENERGY RESULTS}

\begin{figure}[ht!]
\centering
\includegraphics[width=5.0in]{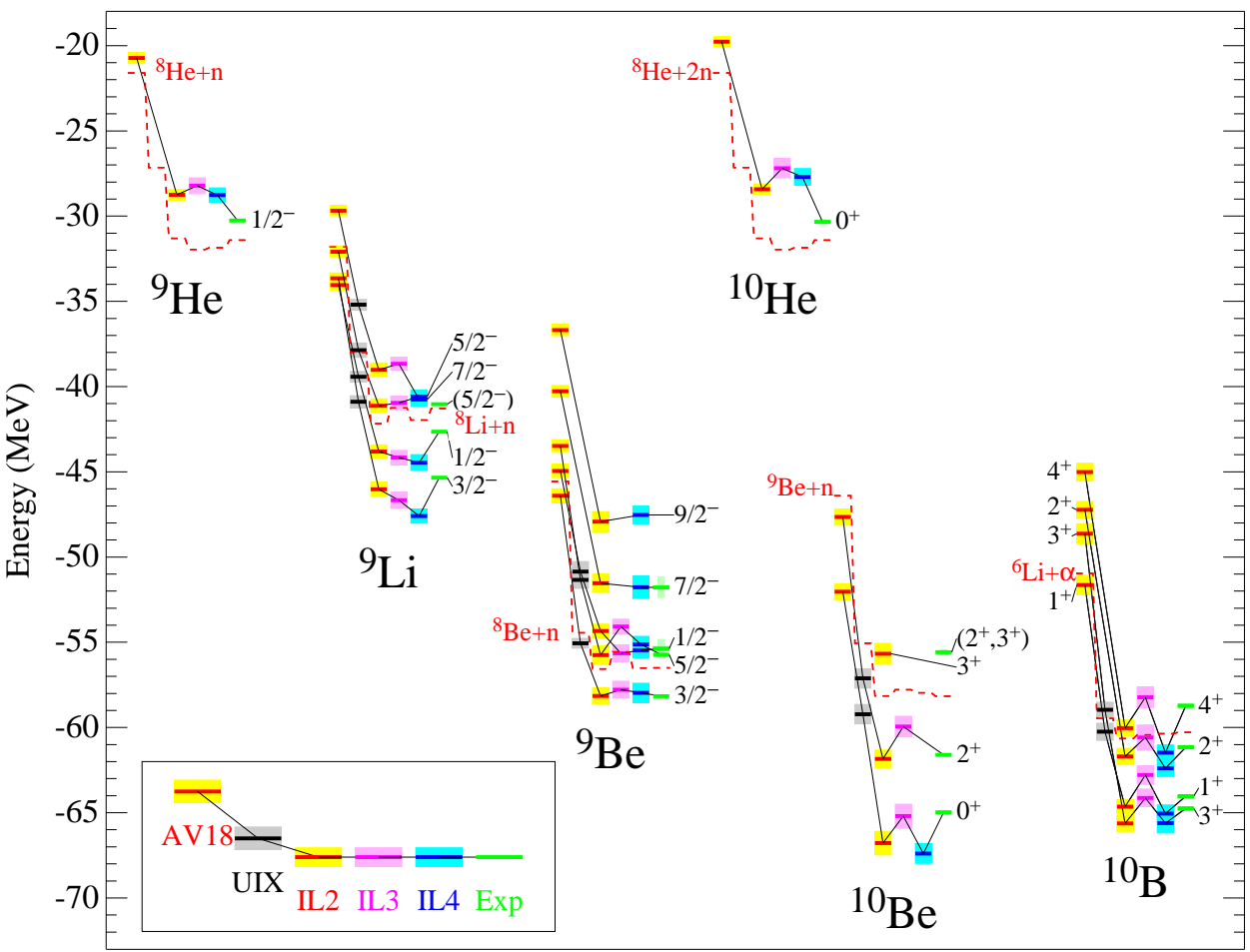}
\caption{GFMC and experimental energies for $A=9,10$ nuclei.  The light
shading shows the Monte Carlo statistical errors or experimental widths.}
\label{fig:GFMC_energy}
\end{figure}

\begin{figure}[ht!]
\centering
\includegraphics[width=5.0in]{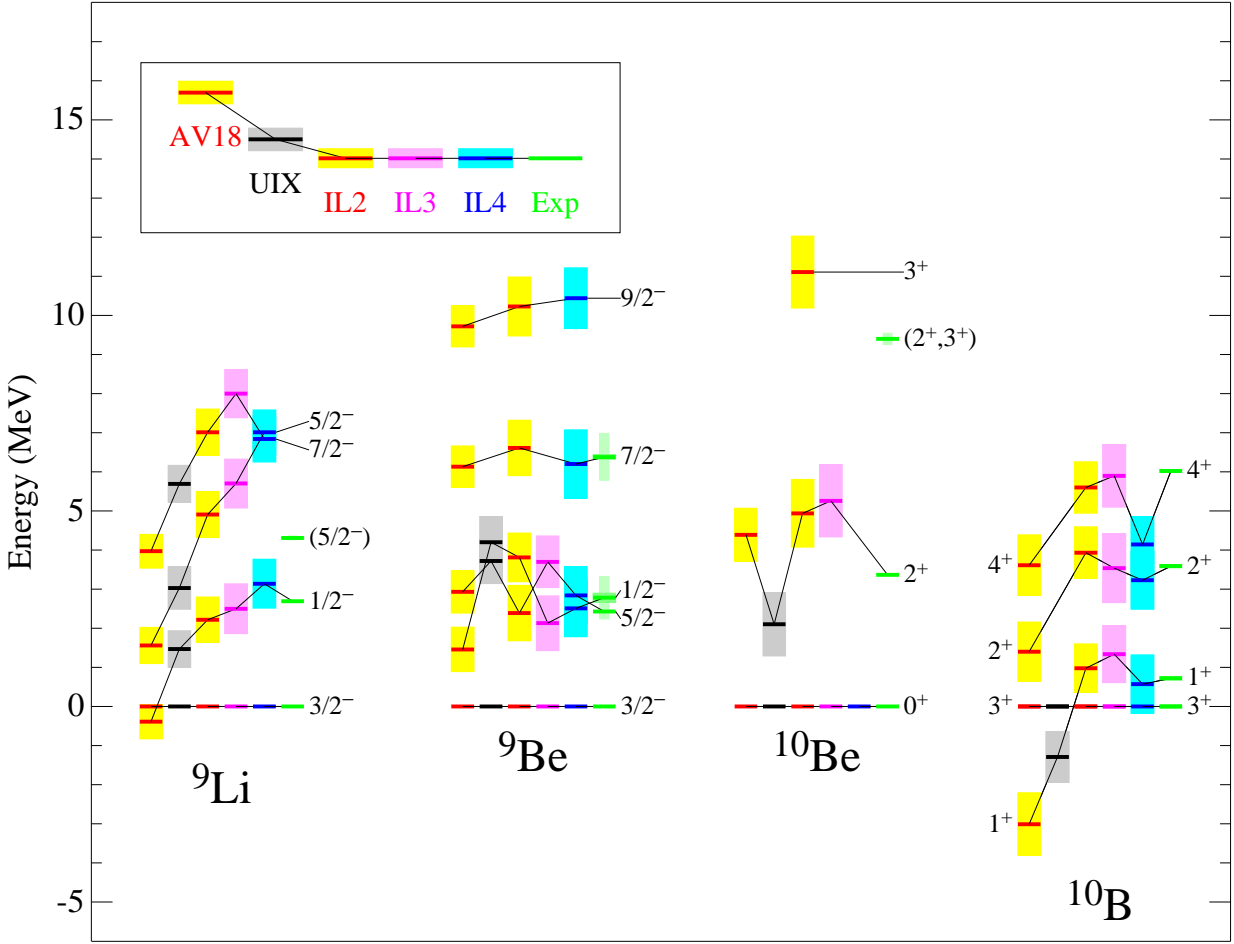}
\caption{GFMC and experimental excitation energies for $A=9,10$ nuclei.}
\label{fig:GFMC_excite}
\end{figure}

In this section we present GFMC energy results in $A=9,10$ nuclei for the 
simplified AV8$^\prime$ $N\!N$ Hamiltonian, the full AV18 interaction, and 
AV18 with each of the $N\!N\!N$ interactions UIX, IL2, IL3, and IL4.
(Results for $A \le 8$ nuclei may be found in Ref.~\cite{PPWC01}.)
The total energies for 18 states are shown in Table~\ref{tab:total}
and in Fig.~\ref{fig:GFMC_energy};
excitation spectra are shown in Fig.~\ref{fig:GFMC_excite}.
Not all states have been calculated with all the Hamiltonians.
Numbers in parentheses are the Monte Carlo statistical errors; in addition
there may be systematic errors from the GFMC algorithm of the order 
of 1--2\% percent as discussed above.

The energies of the particle-stable ground states calculated with just the $N\!N$
interactions are underbound by 8--12 MeV in $A=9$ and 12--16 MeV in $A=10$,
with the AV8$^\prime$ being somewhat more attractive than AV18.
Addition of the older UIX $N\!N\!N$ interaction to AV18 reduces this 
discrepancy significantly to 2--4 MeV in $A=9$ and 4--6 MeV in $A=10$.
However, addition of the IL2-IL4 models to AV18 results in much better
energies, some high and some low, but generally within 2 MeV.
In particular, the AV18/IL2 and AV18/IL3 Hamiltonians come within $\pm 3\%$ 
of the experimental bindings for the ground states of $^9$Li, $^9$Be, 
$^{10}$Be, and $^{10}$B, which is about as good as can be expected, 
given the 1--2\% total errors in the GFMC calculation.

Despite their systematic underbinding, the older Hamiltonians were able to
give the correct ordering of excited states in the $A \le 8$ nuclei, albeit
usually with insufficient splitting between states that are spin-orbit
partners \cite{PPWC01}.
In the $A=9,10$ nuclei this no longer holds true.
The most dramatic case is the question of the proper ground state for $^{10}$B:
AV8$^\prime$, AV18, and AV18/UIX all clearly and wrongly predict the $(1^+;0)$
state to be lower than the $(3^+;0)$ state, while all the AV18/IL2-IL4 models
correctly reproduce the experimental ordering.
This result for the AV18 interaction was first obtained by Navr\'atil who
also finds it for the CD-Bonn interaction~\cite{navratil}.
This incorrect ordering of states by the two-body Hamiltonians also seems to
be present in the case of the $(\case{5}{2}^-;\case{1}{2})$ and 
$(\case{1}{2}^-;\case{1}{2})$ excitations of $^9$Be, and may be a problem for
the proper ground state of $^9$Li, although with the statistical errors of 
the calculation it is not so clear.
The AV18/UIX and AV18/IL2-IL4 models obtain these states in the proper order, 
with one exception, and generally with fairly good spin-orbit splittings.

As noted in Sec. II, two narrow, particle-unstable states have
been observed in $^9$Li without definite spin assignments.
On the basis of the close agreement between the AV18/IL2 energies for the
lowest $(\case{5}{2}^-;\case{1}{2})$ and $(\case{7}{2}^-;\case{1}{2})$ states
and the observed excitations, as illustrated in Fig.~\ref{fig:GFMC_excite},
we think these are in fact the most likely assignments for these states.
Also, our one calculation of the expected $(3^+;1)$ state in $^{10}$Be lines
up with the observed state at 9.4 MeV, suggesting that is its
proper spin assignment.

The current calculations significantly underbind $^{9,10}$He,
even when the AV18/IL2-IL4 models are used.
Because both these nuclei are particle unstable, they
should properly be calculated as resonant states, so the current GFMC
results may be somewhat less reliable than for the other ground states.
However, the experimental widths of these states are not large, and our
calculations for other comparably narrow states have generally been good.
It may be that some significant admixture of $1s_{1/2}$ orbitals from the 
sd-shell should be introduced in the Jastrow trial function, as suggested 
by recent shell-model studies~\cite{OFUBHM01}; we have not yet attempted
such a modification.  In addition, as was discussed in the previous section,
the constrained-path propagation may not be as reliable for these nuclei.
If the present results for $^{9,10}$He are correct, 
then there is a need for further tuning of the
isospin dependence of the $N\!N\!N$ interactions.

In addition to the total energies discussed above, we have calculated
the energy differences between isobaric analog states perturbatively by
evaluating the expectation values of the electromagnetic and strong CSB and 
CD parts of the AV18 Hamiltonian in the wave function of the $T_z=-T$ nucleus.
The CSB and CD terms can induce corresponding changes in the nuclear wave
functions, leading to higher-order perturbative corrections to the splitting
of isospin mass multiplets, but it is difficult for us to estimate these
higher-order effects reliably in either VMC or GFMC calculations.
The energies for a given isomultiplet of states can be expanded as
\begin{equation}
   E_{A,T}(T_z) = \sum_{n\leq 2T} a_n(A,T) Q_n(T,T_z) \ ,
\label{eq:isoexpand}
\end{equation}
where $Q_0=1$, $Q_1=T_z$, and $Q_2=\case{1}{2}(3T_z^2-T^2)$ are
isoscalar, isovector, and isotensor terms~\cite{P60}.
The isovector coefficients $a_1(A,T)$ are shown in Table~\ref{tab:isovec},
while the isotensor coefficients $a_2(A,T)$, are given in 
Table~\ref{tab:isoten}.

The dominant contribution in each case is from the regular Coulomb interaction
between protons, but there are additional contributions of $\approx 5\%$
coming from the other electromagnetic and strong terms.
Because the AV18 and AV18/UIX models underbind these nuclei, the density
distributions tend to be more diffuse and the Coulomb interaction is reduced,
leaving a sizable discrepancy with experiment for the isovector coefficients.
The AV18/IL2-IL4 models give much better binding energies and consequently
larger isovector coefficients which are within 5\% of the experimental values.
However, the isotensor coefficients come out a bit too large.

\section{MOMENTS AND DENSITY DISTRIBUTIONS}

Point proton and neutron rms radii are shown in Tables~\ref{tab:rms-p} and
\ref{tab:rms-n}.
The AV18/IL2-IL4 Hamiltonians give significantly smaller radii than the 
older models, presumably because of their greater binding.
The experimentally measured charge radii~\cite{rms-expt} for the $^9$Be 
and $^{10}$B ground states are in good agreement with the calculations 
from the new models.

Quadrupole moments are shown in Table~\ref{tab:quad}, as evaluated in impulse
approximation.  Experimental values are from Refs.~\cite{AS88,Til01}.
The contribution of two-body charge operators to the quadrupole moment are 
expected to be small, as in the deuteron and $^6$Li~\cite{WSS95,WS98},
because both the isoscalar and isovector charge operators are relativistic 
corrections of order $(v/c)^2$.
Expectation values of the quadrupole operator tend to have a much larger
variance than radii because of the cancellations from the $Y_{20}$ operator.
Given this caveat, the agreement with the measured quadrupole moments is
reasonable.

The impulse approximation magnetic moments are shown in Table~\ref{tab:mag};
experimental values are from Refs.~\cite{AS88,Til01,9CMM,9CMMb}.
Experience in lighter nuclei shows that there are significant two-body 
current contributions to the magnetic operators, particularly the isovector
part.
In the trinucleons, the isoscalar portion is boosted by $+0.034 \mu_N$ or
$\approx 8\%$, while the isovector portion is corrected by $-0.778 \mu_N$
or $\approx 18\%$~\cite{MRS98}.
However, in the isoscalar $^6$Li case, the correction is a tiny +0.003 
$\mu_N$~\cite{WS98}.
On this basis, we expect substantial improvement in the $^9$Li and $^9$C 
results shown in Table~\ref{tab:mag} when two-body contributions to the 
current operator are added, while the already good $^{10}$B results will
not be significantly affected.
However, the reasonable agreement between the IA calculations and experiment
for $^9$Be may not survive.

\begin{figure}[ht!]
\centering
\includegraphics[width=5.0in]{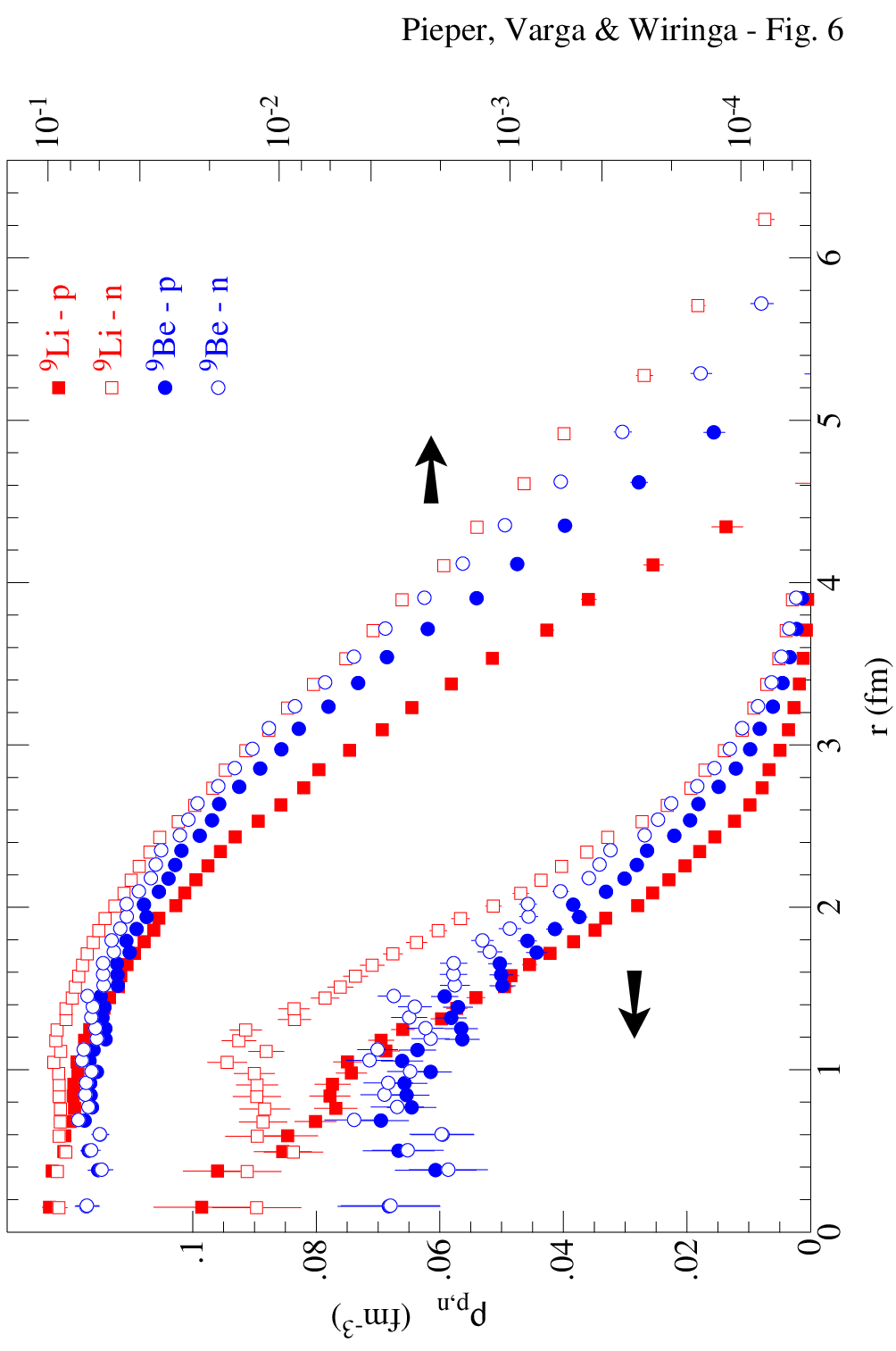}
\caption{Proton and neutron densities for $A=9$ nuclei on both linear
(left) and logarithmic (right) scales.}
\label{fig:dens9}
\end{figure}

\begin{figure}[ht!]
\centering
\includegraphics[width=5.0in]{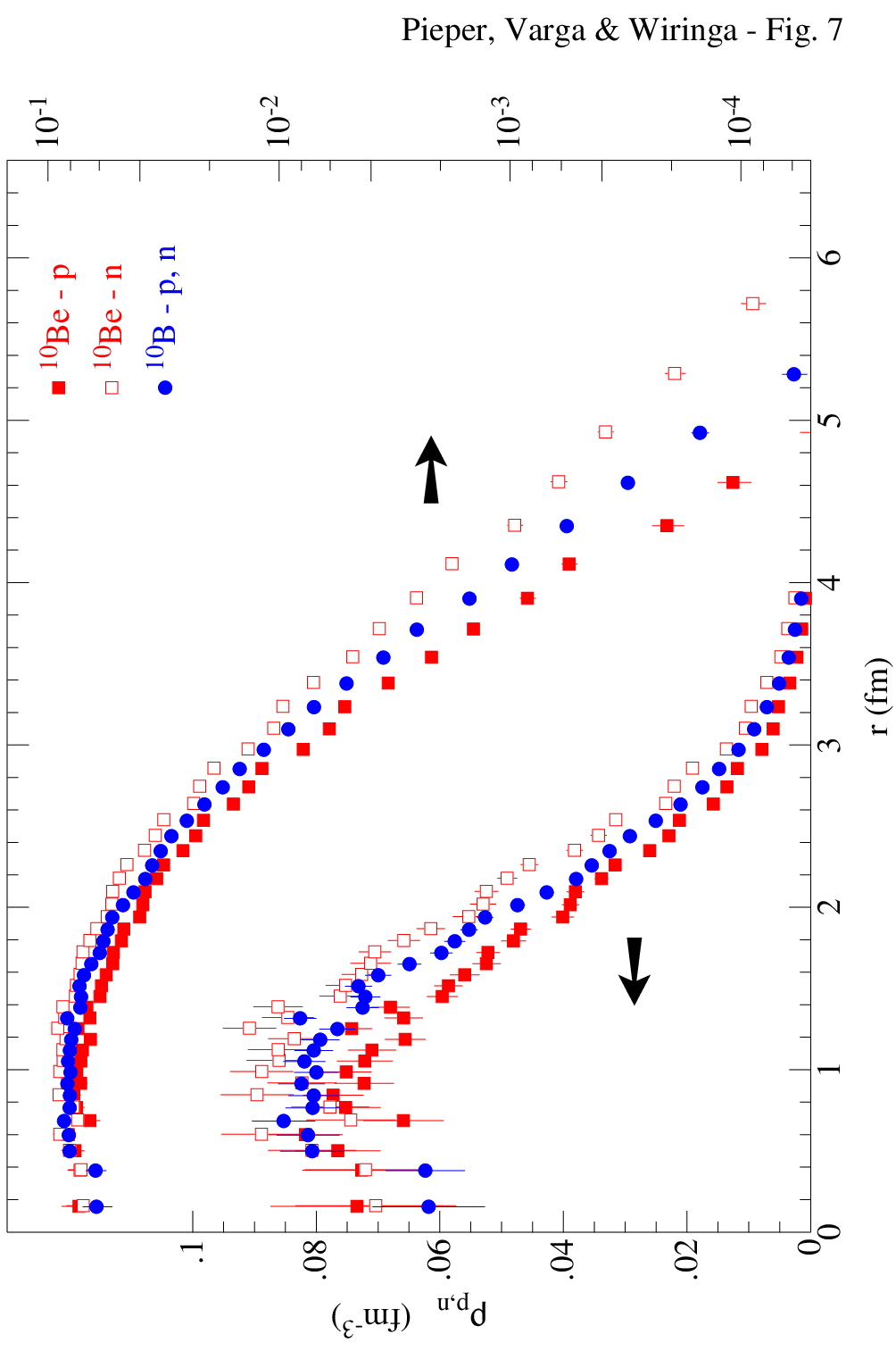}
\caption{Proton and neutron densities for $A=10$ nuclei on both linear
(left) and logarithmic (right) scales.}
\label{fig:dens10}
\end{figure}

The nucleon densities for $^9$Li and $^9$Be are shown in Fig.~\ref{fig:dens9},
on both a linear (left) and logarithmic (right) scale.
The densities are normalized such that the integrated value equals the
appropriate total value of $N$ or $Z$.
These densities are rather similar to those calculated earlier for $^8$Li 
and $^8$Be~\cite{WPCP00} with the former more peaked at the origin,
and the latter showing a broad flat interior.
The protons in $^9$Li cluster at the origin because two out of three are 
confined to the $\alpha$ core, while the neutrons show a substantial skin 
with a hint of a p-shell peak around 1 fm.
The low central densities for both protons and neutrons in $^9$Be are
probably due to the large $2\alpha$ component in its $^8$Be core.
The densities for $^{10}$Be and $^{10}$B are shown in Fig.~\ref{fig:dens10}.
They have somewhat higher central values than in $^9$Be, but are also broader and flatter
than $^9$Li, presumably due to the dominant $^8$Be core.

\section{CONCLUSIONS}

We have made quantum Monte Carlo calculations for the ground states and 
low-lying excited states of $A=9,10$ nuclei, based on realistic two- and 
three-nucleon potentials.
The new Illinois $N\!N\!N$ models do a good job of explaining the binding 
energies of the ten best-known $A=9,10$ states of different quantum numbers,
but the results for $^{9,10}$He may indicate a deficiency in the models.
If we also include the 17 narrow states to which these forces were fit in $A \leq 8$
nuclei~\cite{PPWC01}, we find rms deviations from the experimental energies of 0.6 MeV
for AV18/IL2, 0.8 MeV for AV18/IL3, and 1.0 MeV for AV18/IL4.
This contrasts with rms deviations of 3.2 MeV for AV18/UIX and 9.9 MeV for
AV18 alone.
Thus AV18/IL2 becomes our preferred model Hamiltonian with AV18/IL3 almost
as good, although AV18/IL4 will be somewhat deprecated.

We conclude that a fairly consistent picture of nuclear binding can be 
constructed for $A \leq 10$ nuclei using a single Hamiltonian and a
single computational framework.
This applies also to the energy differences among isobaric multiplets,
which are well-reproduced.
Electromagnetic moments, within the limitation of the impulse approximation,
are in fairly good agreement with experimental data.

Significant challenges for the future will be the computation of second
or higher excited states of the same quantum number, and intruder states
of unnatural parity; both of these types of states first become 
particle-stable in the $A=10$ nuclei.
We also need to evaluate electroweak transition rates in these nuclei, 
as has been done previously in the $A=6,7$ systems~\cite{WS98,SW02}.
The present calculations are at the limit of our current computer resources,
both in CPU time and resident memory.
However, the present codes are in principle capable of computing nuclei
like $^{11}$Be and $^{12}$C; such calculations may become feasible
with the next generation of computers.

\acknowledgments
The authors thank J. Carlson, D. Kurath, J. Millener, V. R. Pandharipande,
and R. Schiavilla for many useful discussions and suggestions.  
The calculations were made possible by generous grants of time on
the IBM SP at the National Energy Research Scientific Computing Center
and the IBM SP, SGI Origin 2000, and Chiba City Linux cluster of the 
Mathematics and Computer Science Division, Argonne National Laboratory.  
This work is supported by the U. S. Department of Energy, Nuclear Physics 
Division, under contract No. W-31-109-ENG-38.

\vfill
\newpage

\begin{table}[ht!]
\caption{Permutation symmetry terms for $LS$-coupling in $A=9,10$ nuclei
and corresponding spin states.}
\label{tab:perms}
\begin{tabular}{rllll}
  $A$  & $[n]  $ & $L    $ & $(T,S)$ & highest symmetry states \\
\colrule
  $9$  & $[41] $ & $1,2,3,4$ & $(\case{1}{2},\case{1}{2})$
       & $^9$Be$(\case{1}{2}^-$--~$\case{9}{2}^-)$ \\
       & $[32] $ & $1,2,3  $ & $(\case{3}{2},\case{1}{2})
                                (\case{1}{2},\case{3}{2})
                                (\case{1}{2},\case{1}{2})     $
       & $^9$Li$(\case{1}{2}^-$--~$\case{7}{2}^-)$ \\
       & $[311]$ & $0,2    $ & $(\case{3}{2},\case{3}{2})
                                (\case{3}{2},\case{1}{2})
                                (\case{1}{2},\case{3}{2})
                                (\case{1}{2},\case{1}{2})     $ \\
       & $[221]$ & $1      $ & $(\case{5}{2},\case{1}{2})
                                (\case{3}{2},\case{3}{2})
                                (\case{3}{2},\case{1}{2})
                                (\case{1}{2},\case{5}{2})
                                (\case{1}{2},\case{3}{2})
                                (\case{1}{2},\case{1}{2})     $
       & $^9$He$(\case{1}{2}^-,\case{3}{2}^-)$ \\
  $10$ & $[42] $ & $0,2^2,3,4$ & $(1,0) (0,1) $
       & $^{10}$Be$(0^+,2^+$--~$4^+)$, $^{10}$B$(1^+$--~$5^+)$ \\
       & $[33] $ & $1,3      $ & $(1,1) (0,0) $
       & $^{10}$Be$(1^+)$                                              \\
       & $[411]$ & $1,3      $ & $(1,1) (0,0) $ \\
       & $[321]$ & $1,2      $ & $(2,1) (2,0) (1,2) (1,1)^2 (1,0) (0,2) (0,1) $
       & $^{10}$Li$(0^+$--~$3^+)$, $^{10}$B$(0^+)$                 \\
       & $[222]$ & $0        $ & $(3,0) (2,1) (1,2) (1,0) (0,3) (0,1) $
       & $^{10}$He$(0^+)$                                              \\
\end{tabular}
\end{table}

\begin{table}[ht!]
\caption{VMC diagonalization for $\beta_{LS[n]}$ components in $^9$Be.}
\begin{tabular}{lrrrrrrrrrr}
$J^{\pi}$       & $^2$P[41] & $^2$D[41] & $^2$F[41] & $^2$G[41] & $^4$P[32] 
                & $^4$D[32] & $^4$F[32] & $^2$P[32] & $^2$D[32] & $^2$F[32] \\
\colrule
$\case{3}{2}^-$ &   0.936   & --0.337   &           &           &   0.035   
                & --0.024   &   0.047   & --0.026   &   0.049   &           \\
$\case{5}{2}^-$ &           &   0.952   &   0.273   &           & --0.064   
                & --0.040   &   0.060   &           &   0.062   & --0.041   \\
$\case{1}{2}^-$ &   0.990   &           &           &           &   0.117   
                & --0.039   &           &   0.002   &           &           \\
$\case{7}{2}^-$ &           &           &   0.868   &   0.488   &           
                &   0.019   & --0.064   &           &           &   0.034   \\
$\case{3}{2}^-$*&   0.356   &   0.921   &           &           &   0.087   
                & --0.098   &   0.052   & --0.050   & --0.003   &           \\
$\case{5}{2}^-$*&           & --0.248   &   0.925   &           &   0.273   
                &   0.007   &   0.045   &           &   0.037   & --0.051   \\
$\case{9}{2}^-$ &           &           &           &   0.978   &           
                &           &   0.207   &           &           &           \\
$\case{1}{2}^-$*& --0.104   &           &           &           &   0.905   
                &   0.286   &           & --0.153   &           &           \\
$\case{7}{2}^-$*&           &           & --0.474   &   0.836   &           
                & --0.260   &   0.009   &           &           &   0.042   \\
\colrule
$J^{\pi}$       & $^4$S[311]& $^4$D[311]& $^2$S[311]& $^2$D[311]
                & $^6$P[221]& $^4$P[221]& $^2$P[221] \\
\colrule
$\case{3}{2}^-$ &   0.046   & --0.025   &           &   0.011   
                &   0.029   & --0.010   &   0.010    \\
$\case{5}{2}^-$ &           &   0.053   &           & --0.005   
                & --0.025   & --0.015   &            \\
$\case{1}{2}^-$ &           & --0.054   & --0.023   &           
                &           & --0.038   &   0.024    \\
$\case{7}{2}^-$ &           &   0.057   &           &           
                & --0.015   &           &            \\
$\case{3}{2}^-$*&   0.046   &   0.035   &           &  -0.000   
                & --0.005   &   0.000   &   0.010    \\
$\case{5}{2}^-$*&           &   0.003   &           & --0.035   
                & --0.017   &   0.021   &            \\
$\case{9}{2}^-$ &           &           &           &
                &           &           &            \\
$\case{1}{2}^-$*&           & --0.112   &   0.216   &           
                &           & --0.077   &   0.005    \\
$\case{7}{2}^-$*&           & --0.080   &           &           
                & --0.015   &           &            \\
\end{tabular}
\label{tab:be9}
\end{table}

\begin{table}[ht!]
\caption{VMC diagonalization for $\beta_{LS[n]}$ components in $^9$Li.}
\begin{tabular}{lrrrrrrrrr}
$J^{\pi}$     & $^2$P[32] & $^2$D[32] & $^2$F[32] & $^4$S[311]& $^4$D[311]
              & $^2$S[311]& $^2$D[311]& $^4$P[221]& $^2$P[221] \\
\colrule
$\case{3}{2}^-$
              &   0.859   & --0.089   &           & --0.446   & --0.070
              &           & --0.198   &   0.107   & --0.004    \\
$\case{1}{2}^-$
              &   0.841   &           &           &           & --0.487
              &   0.219   &           &   0.028   &   0.085    \\
$\case{5}{2}^-$
              &           &   0.968   & --0.234   &           & --0.045
              &           & --0.060   & --0.042   &            \\
$\case{3}{2}^-$*
              & --0.130   &   0.844   &           & --0.489   &   0.016
              &           &   0.160   & --0.017   & --0.075    \\
$\case{7}{2}^-$
              &           &           &   0.826   &           & --0.563
              &           &           &           &            \\
$\case{1}{2}^-$*
              & --0.029   &           &           &           &   0.333
              &   0.934   &           & --0.109   &   0.054    \\
$\case{5}{2}^-$*
              &           & --0.088   & --0.518   &           &   0.821
              &           &   0.163   & --0.154   &            \\
\end{tabular}
\label{tab:li9}
\end{table}

\begin{table}[ht!]
\caption{VMC diagonalization for $\beta_{LS[n]}$ components in $^{10}$B.}
\begin{tabular}{lrrrrrrrrr}
$J^{\pi}$ & $^3$S[42] & $^3$D[42] & $^3$D[42] & $^3$F[42] & $^3$G[42] 
          & $^1$P[33] & $^1$F[33] & $^1$P[411]& $^1$F[411] \\
\colrule
$3^+$     &           &   0.036   &   0.995   & --0.086   &   0.001
          &           &   0.016   &           &   0.003    \\
$1^+$     &   0.889   & --0.225   &   0.368   &           &
          &   0.068   &           & --0.137   &            \\
$1^+$*    &   0.179   &   0.922   &   0.024   &           & 
          & --0.021   &           & --0.342   &            \\
$2^+$     &           &   0.467   &   0.884   &   0.023   &
          &           &           &           &            \\
$3^+$*    &           &   0.917   &   0.368   &   0.016   & --0.047
          &           &   0.021   &           &   0.142    \\
$1^+$**   & --0.042   & --0.014   &   0.998   &           & 
          & --0.025   &           & --0.038   &            \\
$2^+$*    &           & --0.141   &   0.978   & --0.154   &
          &           &           &           &            \\
$4^+$     &           &           &           &   0.901   &   0.433
          &           &           &           &            \\
\end{tabular}
\label{tab:b10}
\end{table}

\begin{table}[ht!]
\caption{VMC diagonalization for $\beta_{LS[n]}$ components in $^{10}$Be.}
\begin{tabular}{lrrrrrrrrr}
$J^{\pi}$ & $^1$S[42] & $^1$D[42] & $^1$D[42] & $^1$F[42] & $^1$G[42]
          & $^3$P[33] & $^3$F[33] & $^3$P[411]& $^3$F[411] \\
\colrule 
$0^+$     &   0.812   &           &           &           &
          &   0.109   &           &   0.573   &            \\
$2^+$     &           &   0.944   & --0.071   &           &
          & --0.050   &   0.005   &   0.288   & --0.138    \\
$2^+$*    &           & --0.035   &   0.998   &           &
          & --0.030   &   0.019   &   0.032   &   0.017    \\
$0^+$*    & --0.462   &           &           &           &
          & --0.492   &           &   0.738   &            \\
$3^+$     &           &           &           &   0.969   &
          &           &   0.102   &           & --0.225    \\
$4^+$     &           &           &           &           &   0.867
          &           & --0.228   &           & --0.443    \\
\end{tabular}
\label{tab:be10}
\end{table}

\begin{table}[ht!]
\caption{Total GFMC energies in MeV.}
\label{tab:total}
\begin{tabular}{lrrrrrrl}
                 & \mbox{AV8$^\prime$}~ & AV18~~  &   UIX~~~   &    IL2~~~  &    IL3~~~  &   IL4~~~   & ~~Expt \\
\hline
$^9$He($\case{1}{2}^-$) & $-$22.5(2) & $-$20.7(3) &            & $-$28.7(3) & $-$28.2(5) & $-$28.8(4) & $-$30.21(8) \\
$^9$Li($\case{3}{2}^-$) & $-$36.6(2) & $-$33.7(3) & $-$40.9(3) & $-$46.0(4) & $-$46.7(5) & $-$47.6(4) & $-$45.34 \\
$^9$Li($\case{1}{2}^-$) & $-$36.8(2) & $-$34.0(3) & $-$39.4(3) & $-$43.8(4) & $-$44.2(4) & $-$44.5(5) & $-$42.65 \\
$^9$Li($\case{5}{2}^-$) & $-$35.0(2) & $-$32.1(3) & $-$37.9(4) & $-$41.1(4) & $-$41.0(4) & $-$40.6(4) & $-$39.96(?) \\
$^9$Li($\case{7}{2}^-$) & $-$32.0(2) & $-$29.7(3) & $-$35.2(3) & $-$39.0(4) & $-$38.7(4) & $-$40.8(4) & $-$38.91(?) \\
$^9$Be($\case{3}{2}^-$) & $-$49.9(2) & $-$46.4(4) & $-$55.1(3) & $-$58.2(5) & $-$57.8(5) & $-$58.0(6) & $-$58.16 \\
$^9$Be($\case{5}{2}^-$) & $-$47.8(2) & $-$43.5(3) & $-$51.3(5) & $-$55.8(5) & $-$54.1(4) & $-$55.1(5) & $-$55.73 \\
$^9$Be($\case{1}{2}^-$) & $-$48.2(2) & $-$45.0(4) & $-$50.9(6) & $-$54.3(4) & $-$55.7(5) & $-$55.5(4) & $-$55.36 \\
$^9$Be($\case{7}{2}^-$) & $-$43.5(2) & $-$40.3(3) &            & $-$51.5(5) &            & $-$51.8(7) & $-$51.78 \\
$^9$Be($\case{9}{2}^-$) & $-$40.1(2) & $-$36.7(3) &            & $-$47.9(6) &            & $-$47.5(5) &          \\
$^{10}$He(0$^+$)        & $-$21.2(2) & $-$19.8(3) &            & $-$28.4(3) & $-$27.2(6) & $-$27.7(5) & $-$30.34 \\
$^{10}$Be(0$^+$)        & $-$56.1(2) & $-$52.0(5) & $-$59.2(6) & $-$66.8(7) & $-$65.2(7) & $-$67.4(6) & $-$64.98 \\
$^{10}$Be(2$^+$)        & $-$51.9(2) & $-$47.7(5) & $-$57.1(6) & $-$61.8(5) & $-$59.9(6) & $-$61.1(6) & $-$61.61 \\
$^{10}$Be(3$^+$)        &            &            &            & $-$55.7(6) &            &            & $-$55.58 \\
$^{10}$B(3$^+$)         & $-$53.2(3) & $-$48.6(6) & $-$59.0(4) & $-$65.6(5) & $-$64.1(5) & $-$65.6(6) & $-$64.75 \\
$^{10}$B(1$^+$)         & $-$55.7(3) & $-$51.6(6) & $-$60.3(5) & $-$64.7(4) & $-$62.8(5) & $-$65.1(5) & $-$64.03 \\
$^{10}$B(2$^+$)         & $-$52.2(2) & $-$47.2(5) &            & $-$61.7(5) & $-$60.6(7) & $-$62.4(5) & $-$61.16 \\
$^{10}$B(4$^+$)         & $-$50.0(3) & $-$45.0(5) &            & $-$60.0(5) & $-$58.2(6) & $-$61.5(4) & $-$58.72 \\
\end{tabular}
\end{table}

\begin{table}[ht!]
\caption{Isovector energies in keV.}
\label{tab:isovec}
\begin{tabular}{lllllll}
            & ~~AV18  & ~~UIX  & ~~IL2    & ~~IL3    & ~~IL4    & EXPT \\
\hline
$^9$Li($\case{3}{2}^-$) & 1900(8)  & 1860(7)  & 2000(7)  & 2041(7)  & 2113(7)  & 2104 \\
$^9$Li($\case{1}{2}^-$) & 1831(8)  & 1804(8)  & 2003(7)  & 1891(7)  & 2047(8)  & 1946 \\
$^9$Be($\case{3}{2}^-$) & 1751(10) & 1738(9)  & 1857(10) & 1833(10) & 1878(14) & 1851 \\
$^9$Be($\case{5}{2}^-$) & 1700(10) & 1704(8)  & 1753(12) & 1754(12) & 1869(13) & 1783 \\
$^{10}$Be(0$^+$)        & 1970(10) & 2115(12) & 2235(12) & 2140(13) & 2307(11) & 2329 \\
$^{10}$Be(2$^+$)        & 2111(11) & 1970(12) & 2189(12) & 2124(11) & 2079(11) & 2322 \\
\end{tabular}
\end{table}

\begin{table}[ht!]
\caption{Isotensor energies in keV.}
\label{tab:isoten}
\begin{tabular}{lllllll}
           & ~AV18 & ~~UIX  & ~~IL2   & ~~IL3   & ~~IL4   & EXPT \\
\hline
$^9$Li($\case{3}{2}^-$) & 182(9)  & 182(9)  & 197(8)  & 202(11) & 204(9)  & 176 \\
$^9$Li($\case{1}{2}^-$) & 172(7)  & 181(8)  & 193(8)  & 181(8)  & 184(10) & 160 \\
$^{10}$Be(0$^+$)        & 257(25) & 205(33) & 250(24) & 284(29) & 314(27) & 241 \\
$^{10}$Be(2$^+$)        & 223(27) & 196(35) & 216(18) & 183(23) & 208(28) & 199 \\
\end{tabular}
\end{table}

\begin{table}[ht!]
\caption{Proton rms radii in fm.}
\label{tab:rms-p} 
\begin{tabular}{lllllll}
           & ~~\mbox{AV8$^\prime$} & ~~UIX  & ~~IL2   & ~~IL3   & ~~IL4   & EXPT \\
\hline
$^9$Li($\case{3}{2}^-$) & 2.19(1) & 2.20(1) & 2.04(1) & 2.02(1) & 1.94(1) &          \\
$^9$Li($\case{1}{2}^-$) & 2.27(1) & 2.23(1) & 2.11(1) & 2.29(1) & 2.07(1) &          \\
$^9$Be($\case{3}{2}^-$) & 2.41(1) & 2.41(1) & 2.38(1) & 2.36(1) & 2.33(1) & 2.40(1)  \\
$^9$Be($\case{5}{2}^-$) & 2.41(1) & 2.41(1) & 2.38(1) & 2.46(1) & 2.27(1) &          \\
$^{10}$Be(0$^+$)        & 2.38(1) & 2.30(1) & 2.21(1) & 2.32(1) & 2.19(1) &          \\
$^{10}$Be(2$^+$)        & 2.33(1) & 2.42(1) & 2.26(1) & 2.29(1) & 2.28(1) &          \\
$^{10}$B(3$^+$)         & 2.40(1) & 2.49(1) & 2.33(1) & 2.38(1) & 2.27(1) & 2.33(12) \\
$^{10}$B(1$^+$)         & 2.45(1) & 2.48(1) & 2.49(1) & 2.52(1) & 2.42(1) &          \\
$^{10}$B(2$^+$)         & 2.46(1) &         & 2.38(1) & 2.59(1) & 2.29(1) &          \\
$^{10}$B(4$^+$)         & 2.42(1) &         & 2.30(0) & 2.41(1) & 2.15(0) &          \\
\end{tabular}
\end{table}

\begin{table}[ht!]
\caption{Neutron rms radii in fm.}
\label{tab:rms-n} 
\begin{tabular}{llllll}
                        & ~~AV18  & ~~UIX   & ~~IL2   & ~~IL3   & ~~IL4  \\
\hline
$^9$Li($\case{3}{2}^-$) & 2.72(1) & 2.76(1) & 2.57(1) & 2.52(1) & 2.39(1) \\
$^9$Li($\case{1}{2}^-$) & 2.84(1) & 2.80(1) & 2.56(1) & 2.76(1) & 2.53(1) \\
$^9$Be($\case{3}{2}^-$) & 2.63(1) & 2.63(1) & 2.55(1) & 2.56(1) & 2.52(1) \\
$^9$Be($\case{5}{2}^-$) & 2.67(1) & 2.65(1) & 2.62(1) & 2.65(1) & 2.48(1) \\
$^{10}$Be(0$^+$)        & 2.76(1) & 2.56(1) & 2.47(1) & 2.59(1) & 2.45(1) \\
$^{10}$Be(2$^+$)        & 2.66(1) & 2.78(1) & 2.55(1) & 2.60(1) & 2.62(1) \\
$^{10}$B(3$^+$)         & 2.40(1) & 2.49(1) & 2.33(1) & 2.38(1) & 2.27(1) \\
$^{10}$B(1$^+$)         & 2.45(1) & 2.48(1) & 2.49(1) & 2.52(1) & 2.42(1) \\
$^{10}$B(2$^+$)         & 2.46(1) &         & 2.38(1) & 2.59(1) & 2.29(1) \\
$^{10}$B(4$^+$)         & 2.42(1) &         & 2.30(0) & 2.41(1) & 2.15(0) \\
\end{tabular}
\end{table}

\begin{table}[ht!]
\caption{Quadrupole moments in fm$^2$.}
\label{tab:quad} 
\begin{tabular}{lrrrrrr}
             & \mbox{AV8$^\prime$}~~ &   UIX~~   &   IL2~~   &   IL3~~   & IL4~~     & EXPT \\
\hline
$^9$Li($\case{3}{2}^-$) &  $-$3.1(1) & $-$2.9(1) & $-$2.7(1) & $-$2.7(1) & $-$2.5(1) & $-$2.7(1) \\
$^9$Be($\case{3}{2}^-$) &     5.0(3) &    7.4(2) &    8.5(3) &    5.7(2) &    6.8(3) & 5.9(1) \\
$^9$Be($\case{5}{2}^-$) &  $-$2.3(2) & $-$2.6(2) & $-$2.9(2) & $-$4.8(2) & $-$4.1(2) &        \\
$^9$C($\case{3}{2}^-$)  &  $-$3.7(3) & $-$5.5(3) & $-$5.1(3) & $-$4.4(3) & $-$5.3(3) &           \\
$^{10}$Be(2$^+$)        & $-$10.1(3) & $-$8.0(4) & $-$5.0(4) & $-$9.6(4) &    7.2(2) &        \\
$^{10}$B(3$^+$)         &     8.7(2) &   12.0(2) &    9.5(2) &    9.4(2) &    8.8(2) & 8.5(1) \\
$^{10}$B(1$^+$)         &     1.2(1) &    3.2(2) &    3.1(1) &    3.2(1) &    2.1(1) &        \\
$^{10}$B(2$^+$)         &  $-$1.9(2) &           & $-$0.6(2) & $-$1.8(3) & $-$1.8(2) &        \\
$^{10}$B(4$^+$)         &     3.9(3) &           &    3.8(2) &    5.4(4) &    3.8(2) &        \\
\end{tabular}
\end{table}

\begin{table}[ht!]
\caption{Magnetic moments in $\mu_N$.}
\label{tab:mag} 
\begin{tabular}{lrrrrrr}
             & \mbox{AV8$^\prime$}~~ &   UIX~~    &   IL2~~    &   IL3~~    & IL4~~      & EXPT \\
\hline
$^9$Li($\case{3}{2}^-$) &    2.91(1) &    2.54(2) &    2.47(2) &    2.54(2) &    2.54(2) &    3.44(0)  \\
$^9$Li($\case{1}{2}^-$) & $-$0.23(2) & $-$0.19(3) & $-$0.29(2) & $-$0.25(3) & $-$0.24(3) &             \\
$^9$Be($\case{3}{2}^-$) & $-$1.35(2) & $-$1.11(1) & $-$1.14(2) & $-$1.03(2) & $-$1.17(3) & $-$1.18(0)  \\
$^9$Be($\case{5}{2}^-$) & $-$0.95(1) & $-$0.96(1) & $-$0.88(1) & $-$0.84(1) & $-$0.85(1) &             \\
$^9$C($\case{3}{2}^-$)  & $-$1.08(3) & $-$0.71(5) & $-$0.63(4) & $-$0.72(4) & $-$0.70(4) & $-$1.39(0)  \\
$^{10}$B(3$^+$)         &    1.82(1) &            &    1.80(1) &            &    1.80(1) &    1.80(0)  \\
$^{10}$B(1$^+$)         &    0.75(1) &    0.78(1) &    0.74(1) &            &    0.77(1) &    0.63(12) \\
\end{tabular}
\end{table}

\vfill
\newpage

\end{document}